\documentclass[10pt]{emulateapj}
\usepackage{apjfonts}
\usepackage{graphicx}
\usepackage{psfig}
\slugcomment{{\sc Submitted to ApJ}}
\makeatletter

\newcommand{\Rmnum}[1]{\expandafter\@slowromancap\romannumeral #1@}
\makeatother
\begin{document}        

\title{Assessing Radiation Pressure as a Feedback Mechanism in Star-Forming Galaxies}
\author{Brett H.~Andrews\altaffilmark{1} \& Todd A.~Thompson\altaffilmark{1,2,3}}
\altaffiltext{1}{Department of Astronomy, 
The Ohio State University, 140 West 18th Avenue, 
Columbus, OH 43210, andrews@astronomy.ohio-state.edu}
\altaffiltext{2}{Center for Cosmology and Astroparticle Physics, 
The Ohio State University, 191 West Woodruff Avenue, 
Columbus, OH 43210.}
\altaffiltext{3}{Alfred P. Sloan Fellow}

\begin{abstract}
Radiation pressure from the absorption and scattering of starlight by dust
grains may be an important feedback mechanism in regulating star-forming
galaxies. We compile data from the literature on star clusters, star-forming
subregions, normal star-forming galaxies, and starbursts to assess the
importance of radiation pressure on dust as a feedback mechanism, by comparing
the luminosity and flux of these systems to their dust Eddington limit.  This
exercise motivates a novel interpretation of the Schmidt Law, the $L_{\rm
IR}$--$L_{\rm CO}^\prime$ correlation, and the $L_{\rm IR}$--$L_{\rm
HCN}^\prime$ correlation.  In particular, the linear $L_{\rm IR}$--$L_{\rm
HCN}^\prime$ correlation is a natural prediction of radiation pressure regulated
star formation.  Overall, we find that the Eddington limit sets a hard upper
bound to the luminosity of any star-forming region.  Importantly, however, many
normal star-forming galaxies have luminosities significantly below the Eddington
limit.  We explore several explanations for this discrepancy, especially the
role of ``intermittency'' in normal spirals---the tendency for only a small
number of subregions within a galaxy to be actively forming stars at any moment
because of the time-dependence of the feedback process and the luminosity
evolution of the stellar population.  If radiation pressure regulates star
formation in dense gas, then the gas depletion timescale is 6 Myr, in good
agreement with observations of the densest starbursts.  Finally, we highlight
the importance of observational uncertainties---namely, the dust-to-gas ratio
and the CO-to-H$_2$ and HCN-to-H$_2$ conversion factors---that must be
understood before a definitive assessment of radiation pressure as a feedback
mechanism in star-forming galaxies.
\end{abstract}

\keywords{galaxies: general, evolution, ISM, stellar content, starburst ---
stars: formation}
%\baselinestretch — \linespread{2}

\section{Introduction}
Understanding global star formation is crucial in understanding galaxy evolution
and the assembly of the $z=0$ stellar population over cosmic time.  Observations
indicate that only a few percent of the available gas reservoir in galaxies is
converted into stars per local free-fall time (Kennicutt 1998; Krumholz \& Tan
2007).  In addition, models of the interstellar medium (ISM) suggest that energy
and momentum injected by massive stars could act as a feedback loop by driving
supersonic turbulence, which would cause most of the gas to be insufficiently
dense to collapse, rendering star formation inefficient (Krumholz \& McKee
2005).  However, the interaction between star formation and the ISM is not
well understood, and a mechanism for the regulation of star formation across the
large dynamic range of star-forming environments has not yet been conclusively
identified.  Proposed mechanisms include supernova explosions, expanding HII
regions, stellar winds, cosmic rays, magnetic fields, and radiation pressure on
dust (McKee \& Ostriker 1977; Matzner 2002; Cunningham 2008; Chevalier \&
Fransson 1984; Socrates et al.~2008; Kim 2003; Scoville et al.~2001; Scoville
2003; Thompson, Quataert, \& Murray 2005, hereafter TQM; Krumholz \& Matzner
2009, hereafter KM09; Murray, Quataert, \& Thompson 2010, hereafter MQT; Draine
2010; Hopkins et al.~2010).\footnote{ISM turbulence driven by non-stellar
processes, such as disk instabilities, has also been proposed (Sellwood \&
Balbus 1999; Wada et al.~2002; Piontek \& Ostriker 2004, 2007).}

In the case of radiation pressure on dust, UV and optical radiation from OB
stars is absorbed and scattered by dust grains and subsequently re-radiated as
IR radiation. The dust grains are coupled to the gas of the ISM through
collisions and magnetic fields, so radiation pressure on the dust exerts a force
on the gas as well (O'dell et al.~1967; Ferrara 1993; Laor \& Draine 1993;
Murray, Quataert, \& Thompson 2005). On galaxy scales, TQM showed that radiation
pressure could constitute the majority of the vertical pressure support in dense
starburst galaxies such as ultra-luminous infrared galaxies (ULIRGs). Likewise,
models of giant molecular cloud (GMC) disruption predict that radiation pressure
is the dominant feedback mechanism regulating star formation in the birth of
massive star clusters (MQT; KM09). In this picture, gas in a marginally-stable
galactic disk collapses to form a GMC and a central compact star cluster.  When
the stellar mass and luminosity of the cluster exceed the Eddington limit for
dust, the overlying gas reservoir is expelled.  Thus, the final stellar mass in
individual star clusters is regulated by the dust Eddington limit.  The centers
of ULIRGs and GMC cores are optically thick to both UV and the re-radiated IR
photons, which make them ideal candidates for radiation pressure support since
essentially all of the momentum from the starlight is efficiently transferred to
the gas.  Recent observations indicate that the most luminous GMCs in the Milky
Way are disrupted by radiation pressure (Murray 2010).

In this paper, we critically assess the theory of radiation pressure regulated
star formation by comparing the picture developed by TQM, MQT, and KM09 with the
available observations of star-forming galaxies ranging from dense individual
star clusters and GMCs, to normal spiral galaxies and starbursts.  In \S
\ref{sec:theory} we describe the current model of radiation pressure feedback.
We emphasize the deviations from the simplest version of the dust Eddington
limit in assessing radiation pressure regulated feedback that arise from
ambiguities in the value of the flux-mean dust opacity, and the tendency for
low-density galaxies to have highly intermittent knots or hotspots of star
formation across their disks.  In \S \ref{sec:results}, we compare data from the
literature to models of radiative feedback.  In \S \ref{sec:discussion}, we
discuss our conclusions, the major observational and theoretical uncertainties
in our analysis, and the implications of our results.

\section{Theoretical Elements}
\label{sec:theory}
The statement that radiation pressure may be an important feedback mechanism in
galaxies is equivalent to the statement that galaxies as a whole or the
star-forming subregions within them approach or exceed the Eddington limit for
dust,
\begin{equation}
 F_{\rm Edd} = \frac{4 \pi G c \Sigma}{\kappa_{\rm F}},
\label{eddf}
\end{equation}
where $F_{\rm Edd}$ is the Eddington flux, $\Sigma$ is the surface density of
the dominant component of gravitational potential in the star-forming region,
and $\kappa_{\rm F}$ is the flux-mean opacity. The overall picture is that
star-forming regions meet the Eddington limit and self-regulate in analogy with
an individual massive star (TQM; see also Scoville et al.~2001; Scoville 2003;
MQT; KM09). We would thus naively expect to test the theory of radiation
pressure regulated star formation by taking the ratio of the observed flux
($F_{\rm obs}$) to $F_{\rm Edd}$. However, a direct comparison between the
simple theoretical expectation
\begin{equation}
F_{\rm obs}/F_{\rm Edd}\rightarrow1
\label{ratio}
\end{equation}
and the observations is complicated by both theoretical and observational
uncertainties. For example, although gas is expected to be the dominant mass
component in and around massive star clusters in formation, it is unclear how
best to estimate $\Sigma$ in equation (\ref{eddf}) for unresolved galaxies or
unresolved star-forming subregions.  Below, we consider both CO and HCN emission
(see \S \ref{sec:results}; Figures \ref{fig:lirlco} \& \ref{fig:lirlhcn}), but
the conversion from the luminosity in either of these molecular gas tracers to
gas mass where the stars are forming, is highly uncertain.  Another uncertainty
is the coupling of the radiation field to the gas, which is complicated due to
both the non-gray nature of the dust opacity and the clumpiness of the gas on
all scales (see \S\ref{sec:radpressure} below). Finally, there is an additional
complication not readily apparent from the time-independent statement of
equation (\ref{ratio}): the star formation rate (SFR) across the face of a large
spiral galaxy is highly intermittent so that only a small number of subregions
are bright at any time.  As discussed by MQT and in detail below
(\S\ref{sec:intermittency}), this {\it intermittency} can cause normal
star-forming galaxies to be appear significantly sub-Eddington ($F_{\rm
obs}/F_{\rm Edd}\ll 1$) when only their average properties are considered, but
much closer to Eddington when a model is used to take this effect into account.

\subsection{The Radiation Pressure Force}
\label{sec:radpressure}
The coupling between radiation and gas in star-forming environments is complex
primarily because the flux-mean opacity $\kappa_{\rm F}$ in equation
(\ref{eddf}) has a full range of more than 3 dex, depending on whether the SED of
the system considered is dominated by UV or FIR light. However, there are two
distinct regimes: optically thick to UV but thin to the re-radiated FIR and
optically thick to FIR. We call these the ``single-scattering'' and ``optically
thick'' limits, respectively.

\subsubsection{Single-scattering Limit}
\label{sec:sslimit}
Regions in the single-scattering limit are optically thick to the UV but
optically thin to the FIR ($\tau_{\rm FIR}\sim\kappa_{\rm FIR}\Sigma_{\rm
g}/2$). This limit applies over a wide range in surface density:
\begin{equation}
 \Sigma_{\rm g} \lesssim 5000 {\rm \; M_{\odot} \; pc^{-2}} \; \kappa_{2}^{-1}
 \, f_{\rm dg, \, 150}^{-1},
 \label{eqn:sslimit}
\end{equation}
where $\kappa_{\rm FIR} = \kappa_{2} f_{\rm dg, \, 150}$ is the Rosseland-mean
dust opacity with $\kappa_{2} = \kappa/(2 \; {\rm cm^{2} \; g^{-1}})$ (see \S
\ref{sec:optthick}) and $f_{\rm dg, \, 150}=f_{\rm dg} \times 150$ is the
dust-to-gas ratio. In the single-scattering limit, UV photons are absorbed once
and then re-radiated as FIR photons, which free-stream out of the
medium.\footnote{Galaxies with surface densities less than $\sim$5 $\rm M_\odot
\, pc^{-2}$ will be optically thin with respect to dust.  Below this limit, the
ionization of neutral hydrogen will becomes the dominant source of opacity. The
large cross-section ($\sigma_{\rm HI} \approx 6.3 \times 10^{-18} \, \rm cm^2$
per H atom) implies an incredibly small surface density ($\Sigma_{\rm g} \gtrsim
10^{-3} \, \rm M_\odot \, pc^{-2}$) is required for the medium to be optically
thick to ionizing photons.  These ionizing photons transfer momentum directly to
the gas on the same order as the momentum transfer due to the single-scattering
limit for dust.  Thus, we encompass this limit and the single-scattering limit
for dust under the same heading.} Since the column-averaged flux-mean optical
depth in this limit is always equal to unity, the flux-mean opacity for the
single-scattering limit is $\sim$$2/\Sigma_{\rm g}$. The Eddington flux is then
\begin{equation}
 F_{\rm Edd}^{\rm s} \sim 10^{8} \; {\rm L_{\odot} \; kpc^{-2}} \;
 \left(\frac{\Sigma_{\rm g}}{10 \; {\rm M_{\odot}\;pc^{-2}}} \right)^2 \, f_{\rm
 gas}^{-1},
\end{equation}
where $f_{\rm gas} = \Sigma_{\rm g}/\Sigma_{\rm tot}$ is the gas fraction and
$\Sigma_{\rm tot} \equiv \Sigma_{\rm g}+0.1\Sigma_{\star}$ (Wong \& Blitz 2002).
The wide range of column densities over which this limit is applicable implies
that the average medium of most star-forming galaxies, some starbursts, and the
GMCs that constitute them is single-scattering.

\subsubsection{Optically Thick Limit}
\label{sec:optthick}
Dense starbursts and GMCs can reach the high gas surface densities $\Sigma_{\rm
g} \gtrsim 5000 {\rm \; M_{\odot} \; pc^{-2}} \; \kappa_{2}^{-1} \, f_{\rm dg,
\, 150}^{-1}$ necessary to become optically thick to FIR photons ($\tau_{\rm
FIR} \gtrsim 1$). In this case $P_{\rm rad} \sim \tau_{\rm FIR} F/c$, and
$\kappa_{\rm FIR}$ depends on temperature (Bell \& Lin 1994; Semenov et
al.~2003). The functional form of $\kappa_{\rm FIR}$ naturally leads to two
regimes: ``warm'' ($T < 200 \; \rm K$) and ``hot'' ($200 \; {\rm K} < T < T_{\rm
sub}$, where $T_{\rm sub} \sim 1500 \rm \; K$ is the dust sublimation
temperature). For typical numbers, the central temperature of a massive, compact
star cluster is
\begin{eqnarray}
 T^4 \sim \tau T_{\rm eff}^4 \sim \frac{\kappa_{\rm FIR} \Sigma}{2}
\frac{F}{\sigma_{\rm SB}} \sim \frac{\kappa_{\rm FIR} M_{\rm g}}{8 \pi R^2}
\frac{M_\star \Psi}{4 \pi R^2 \sigma_{\rm SB}} \nonumber \\ T \sim 290 \; {\rm
K} \; \kappa_{10}^{1/4} \; \Psi_{3000}^{1/4} \; M_{\rm g,\,6}^{1/4} \;
M_{\star,\,5}^{1/4} \; R_{\rm pc}^{-1},
\end{eqnarray}
where $T_{\rm eff}$ is the effective temperature, $\kappa_{10}=\kappa/(10 \; \rm
cm^2 \; g^{-1})$, $\Psi_{3000}=3000 \; \rm ergs \; s^{-1} \; g^{-1}$ is the
light-to-mass ratio of a zero age main sequence stellar population, $M_{\rm
g,\,6}=M_{\rm g}/(10^6 \; \rm M_\odot)$, and $M_{\star,\,5}=M_{\star}/(10^5 \;
\rm M_\odot)$.

\paragraph{Warm Starbursts}
For $T < 200 \; \rm K$, the Rosseland mean opacity increases as $\kappa_{\rm
FIR}(T) \approx \kappa_o T^2$, where $\kappa_o \approx 2 \times 10^{-4}\; {\rm
cm^{2} \; g^{-1} \; K^{-2}} \; f_{\rm dg, \, 150}$. In this case,
\begin{equation}
 F_{\rm Edd}  \approx  \left(\frac{3\pi Gc\sigma_{\rm SB}}
{\kappa_o^2 \,f_{\rm dg, \, 150}^2 \,f_{\rm gas} }\right)^{1/2} 
\hspace*{-.3cm} \sim 10^{13} \; {\rm L_{\odot} \, kpc^{-2}} \, f_{\rm
gas}^{-1/2} \, f_{\rm dg, \, 150}^{-1}.
\end{equation}
Remarkably, the flux necessary to support the medium is independent of $\Sigma$
(TQM).

\paragraph{Hot Starbursts}
Intense, compact starbursts may have central temperatures greater than 200
K. The corresponding opacity is roughly constant with temperature: $\kappa_{\rm
FIR}(T) \approx 5$--$10 \; {\rm cm^{2} \; g^{-1}} \; f_{\rm dg, \, 150}$ for
temperatures $200 \; {\rm K} \lesssim T \lesssim T_{\rm sub}$. For typical
numbers,
\begin{equation}
 F_{\rm Edd}^{\rm thick} \sim 10^{15} \; {\rm L_{\odot}\;kpc^{-2}} \; \left
 (\frac{\Sigma_{\rm g}}{10^6 \; {\rm M_{\odot}\;pc^{-2}}} \right ) \; f_{\rm
 gas}^{-1/2} \, f_{\rm dg, \, 150}^{-1}.
\end{equation}
The high surface densities necessary to enter this regime may only be attained
in the pc-scale star formation thought to attend the fueling of bright AGN
(Sirko \& Goodman 2003; TQM; Levin 2007).

\subsection{GMC Evolution \& Intermittency}
\label{sec:intermittency}
In order to gauge the importance of GMC evolution and intermittency, we adopt
the simple picture presented by MQT that marginally stable ($Q \approx 1$) disks
fragment into sub-units on the gas disk scale height ($h$) to form GMCs.  An
individual star cluster is born, reaches the critical Eddington luminosity
threshold, and then expels the overlying gas.  Importantly, the timescale for
collapse and expansion of the GMC is the disk dynamical timescale, $t_{\rm
dyn}$, which can be much longer than the characteristic timescale for the
stellar population to decrease in total luminosity, the main-sequence lifetime
of massive stars, $t_{\rm MS}\sim4\times10^6$\,yr.  In this picture, a
low-density star-forming galaxy with radius $r$ should have $\sim \! (r/h)^2$
sub-units, but only a small fraction $\xi$ (the ``intermittency factor'') should
be bright at any one time. If each subregion reaches the Eddington luminosity
for a time $t_{\rm MS}$ and is then dark, and then if a large number of
subregions are averaged, one expects
\begin{equation}
 \xi \equiv \frac{N_{\rm on}}{N_{\rm tot}} \sim \frac{L_{\rm obs}}{L_{\rm Edd}}
 \sim \frac{t_{\rm MS}}{2t_{\rm dyn}+t_{\rm MS}},
 \label{eqn:int}
\end{equation}
where 
\begin{eqnarray}
 t_{\rm dyn} &\sim& \left(\frac{3 \pi (2h)}{32 G \Sigma_{\rm tot}}\right)^{1/2}
 \nonumber \\
\hspace*{-.3cm}&\sim& 3.5\times10^7\,{\rm yr}\,\,\,h_{100}^{1/2}f_{\rm
gas}^{1/2} \left(\frac{10\,\,{\rm M_\odot\,\,pc^{-2}}}{\Sigma_{\rm
g}}\right)^{1/2},
\label{eqn:tdyn}
\end{eqnarray}
$h_{100} = h/(100 \; \rm pc)$, $N_{\rm on}$ is the number of sub-units that are
``on,'' and $N_{\rm tot}$ is the total number of sub-units. For example, the
normal star-forming galaxy M51 has a observed bolometric luminosity ($L_{\rm
obs} = 0.2 L_{\rm Edd}$) that is a factor of $\sim$4 larger than its
intermittency-corrected Eddington luminosity ($L_{\rm Edd}^{\rm int} = 0.05
L_{\rm Edd}$). Although the approximation that the stellar population is bright
for a time $t_{\rm MS}$ and then dark is crude, the parameter $\xi$ gives us a
way to judge the importance of intermittency in normal star-forming galaxies.

Note that for higher densities, $t_{\rm dyn}$ decreases and $\xi \rightarrow 1$
at a critical surface density
\begin{equation}
 \Sigma_{\rm crit} \sim \frac{3 \pi (2h)}{32 G (0.5t_{\rm MS})^2} \sim 3 \times
 10^3 \; {\rm M_\odot \; pc^{-2}} \; h_{100},
 \label{eqn:sigcrit}
\end{equation}
which corresponds with a critical midplane pressure $P_{\rm crit} \sim 5 \times
10^{-8} \; {\rm ergs \; cm^{-3}} \; h_{100}^2$ (see eq. \ref{eqn:pradpmid}).
For $\Sigma_{\rm tot}>\Sigma_{\rm crit}$, massive stars live longer than the
time required to disrupt the parent sub-unit ($t_{\rm MS} > t_{\rm dyn}$).  MQT
argue that in this regime the massive stars continue to drive turbulence in the
gas and maintain hydrostatic equilibrium in a statistical sense until $t_{\rm
MS}$, when the process then repeats until gas exhaustion.

Several additional elements of GMC evolution are important in judging whether or
not the star formation of galaxies is regulated by radiation pressure on dust.
First, the GMCs collapse from regions of size $h^2$, with total mass
$\Sigma_{\rm g}\pi h^2$, and to a size $R_{\rm GMC}=h/\phi$.  This implies that
the surface density of individual GMCs is $\Sigma_{\rm GMC}\sim\phi^2\Sigma_{\rm
g}$, where $\phi$ can be $\sim$2--5 in the Galaxy (MQT).  For example, taking
$\epsilon_{\rm GMC}=M_\star/M_{\rm GMC}$ ($\propto \Sigma_{\rm g}$ in the
single-scattering limit), $\Psi_{3000}$, and assuming $\kappa_{\rm
FIR}\sim\kappa_o T^2$ (appropriate for $\rm T \lesssim 200 \, K$), one finds
that the required gas surface density for a GMC to be optically thick to the FIR
is $\Sigma_{\rm GMC}^{\tau_{\rm FIR}=1}\sim7000\epsilon_{\rm
GMC,\,0.1}^{-1/3}$\,M$_\odot$ pc$^{-2}$.  This GMC gas surface density would
correspond to an average gas surface density for the galaxy that is $\phi^2$
times smaller ($\Sigma_{\rm g}\sim350$--$1800$\,M$_\odot$ pc$^{-2}$).  Thus, the
medium surrounding the central star cluster may be optically thick to the FIR
even if the average gas surface density of the galaxy is less than the naive
estimate given in \S \ref{sec:optthick}.

At surface densities in excess of $\tau_{\rm FIR}=1$ for the GMCs, the models of
MQT rely on the fact that the medium is in fact optically-thick to FIR
radiation.  This is in sharp contrast to the work of KM09, where they argue that
the effective optical depth is always $\sim$1 because instabilities allow the
radiation to leak out of otherwise optically-thick media.  MQT argue that the
effective optical depth must be much larger than unity in the GMCs of dense
starbursts and ULIRGs for radiation pressure to be effective as a feedback
mechanism.  In addition, they show that the effective momentum coupling between
the radiation and the gas can exceed the naive estimate based on a disk-averaged
$\tau_{\rm FIR}$ by a factor of a few in systems as dense as the putative GMCs
in Arp 220 because of the time-dependence of the GMC disruption process (see \S
\ref{sec:tauFIR}).

\section{Results}
\label{sec:results}
%%%%%%%%%%%%%%%%%%%%%%%%%%%%%%%%%%%%%%%%%%%%%%%%%%%%%%%%%%%%%%%%%%%%%%%%%%%%
\begin{figure*}
\centerline{\psfig{file=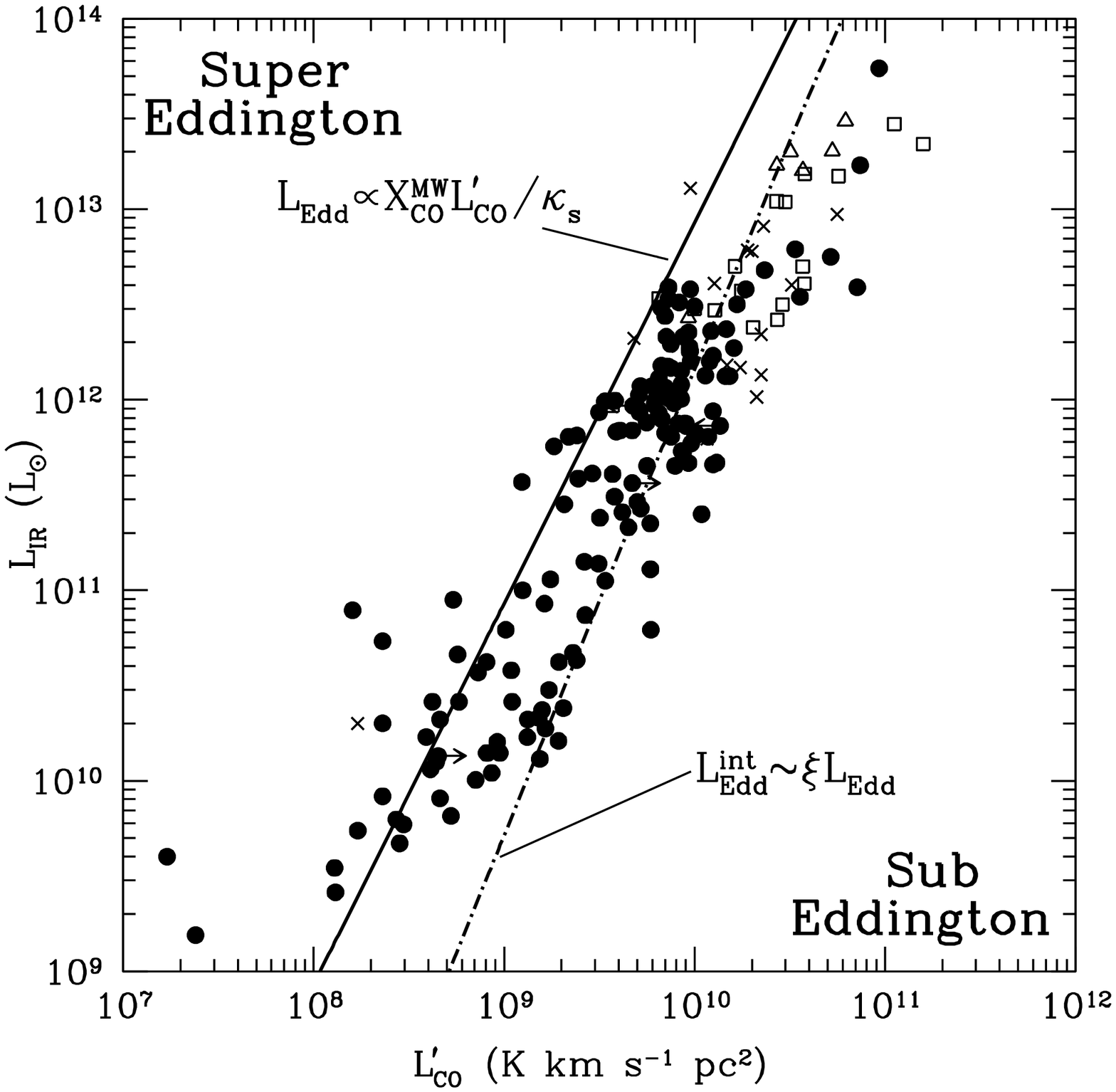,width=9cm}\psfig{file=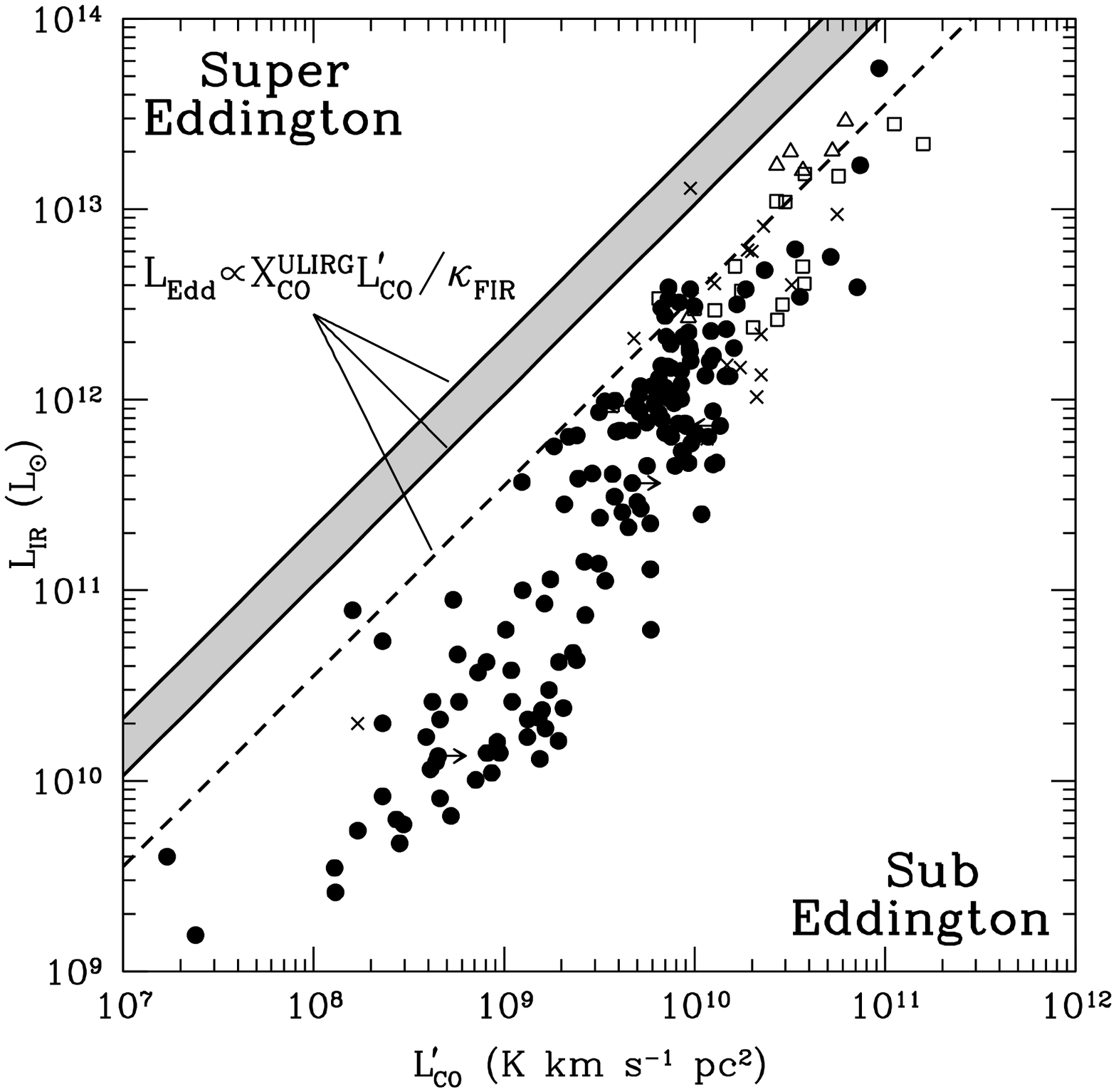,width=9cm}}
\caption{IR luminosity as a function of CO line luminosity.  The different
symbols correspond to different rotational transitions of CO: solid circles
({\it J} = 1--0), crosses ({\it J} = 2--1), squares ({\it J} = 3--2), and
triangles ({\it J} = 4--3). The data are the same in both panels.  The lines in
the left panel correspond to the single-scattering Eddington limit (solid line;
assuming $h = 100 \, \rm pc$ and $r = 10 \, \rm kpc$; \S \ref{sec:sslimit}) and
the single-scattering Eddington limit accounting for intermittency (dot-dashed
line; eq.~\ref{eqn:int}).  The lines in the right panel show the optically thick
Eddington limit for our preferred value of the Rosseland-mean opacity (shaded
region; $\kappa_{\rm FIR}=5$--$10 \; {\rm cm^2 \; g^{-1}} \; f_{\rm dg,150}$)
and for an enhanced dust-to-gas ratio (dashed line; $\kappa_{\rm FIR}=30 \; {\rm
cm^2 \; g^{-1}} \; f_{\rm dg,50}$). Note that no galaxies are significantly
super- or sub-Eddington. We emphasize that it is not possible to determine which
limit is applicable without knowing a surface density, so dense star-forming
regions can be optically thick at $L_{\rm CO}^\prime \lesssim 10^9 \; \rm
K\;km\;s^{-1}\;pc^{2}$ and approach the optically thick Eddington limit (see \S
\ref{sec:radpressure}).  The single-scattering Eddington limit was calculated by
adopting the Galactic CO-to-H$_2$ conversion factor $X_{\rm CO}^{\rm MW}=4.4 \rm
\; M_{\odot} (K\;km\;s^{-1} \;pc^2)^{-1}$. The optically thick Eddington limit
was calculated by adopting the ULIRG CO-to-H$_2$ conversion factor $X_{\rm
CO}^{\rm ULIRG}=0.8\rm \; M_{\odot} (K\;km\;s^{-1} \;pc^2)^{-1}$.
\label{fig:lirlco}}
\end{figure*}
%%%%%%%%%%%%%%%%%%%%%%%%%%%%%%%%%%%%%%%%%%%%%%%%%%%%%%%%%%%%%%%%%%%%%%%%%%%%
%%%%%%%%%%%%%%%%%%%%%%%%%%%%%%%%%%%%%%%%%%%%%%%%%%%%%%%%%%%%%%%%%%%%%%%%%%%%
\begin{figure}
\centerline{\psfig{file=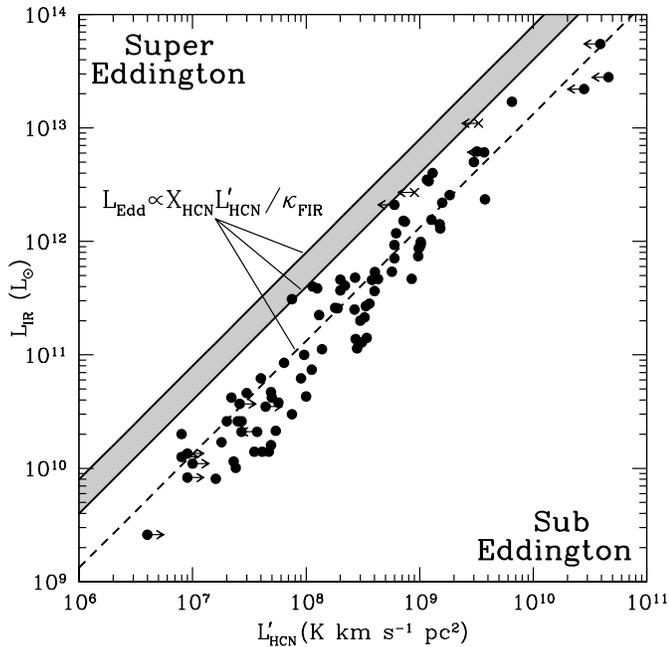,width=9cm}}
\caption{IR luminosity as a function of HCN line luminosity. The different
symbols correspond to different rotational transitions of HCN: solid circles
({\it J} = 1--0) and crosses ({\it J} = 2--1).  We show the optically thick
Eddington limit for our preferred value of the Rosseland-mean opacity (shaded
region; $\kappa_{\rm FIR} = 5$--$10 \; {\rm cm^2 \; g^{-1}} \; f_{\rm
dg,150}$). The dashed line shows the effect of a factor of 3 increase in the
dust-to-gas ratio for the optically thick Eddington limit ($\kappa_{\rm FIR} =
30 \; {\rm cm^2 \; g^{-1}} \; f_{\rm dg,50}$). Note that all galaxies are within
$\sim$1 dex of the optically thick Eddington limit, which suggests that
radiation pressure may regulate star formation.  The optically thick Eddington
limit was calculated by adopting the HCN-to-H$_2$ conversion factor $X_{\rm HCN}
= 3\rm \; M_{\odot} (K\;km\;s^{-1} \;pc^2)^{-1}$.
\label{fig:lirlhcn}}
\end{figure}
%%%%%%%%%%%%%%%%%%%%%%%%%%%%%%%%%%%%%%%%%%%%%%%%%%%%%%%%%%%%%%%%%%%%%%%%%%%%
%%%%%%%%%%%%%%%%%%%%%%%%%%%%%%%%%%%%%%%%%%%%%%%%%%%%%%%%%%%%%%%%%%%%%%%%%%%%
\begin{figure*}
\centerline{\psfig{file=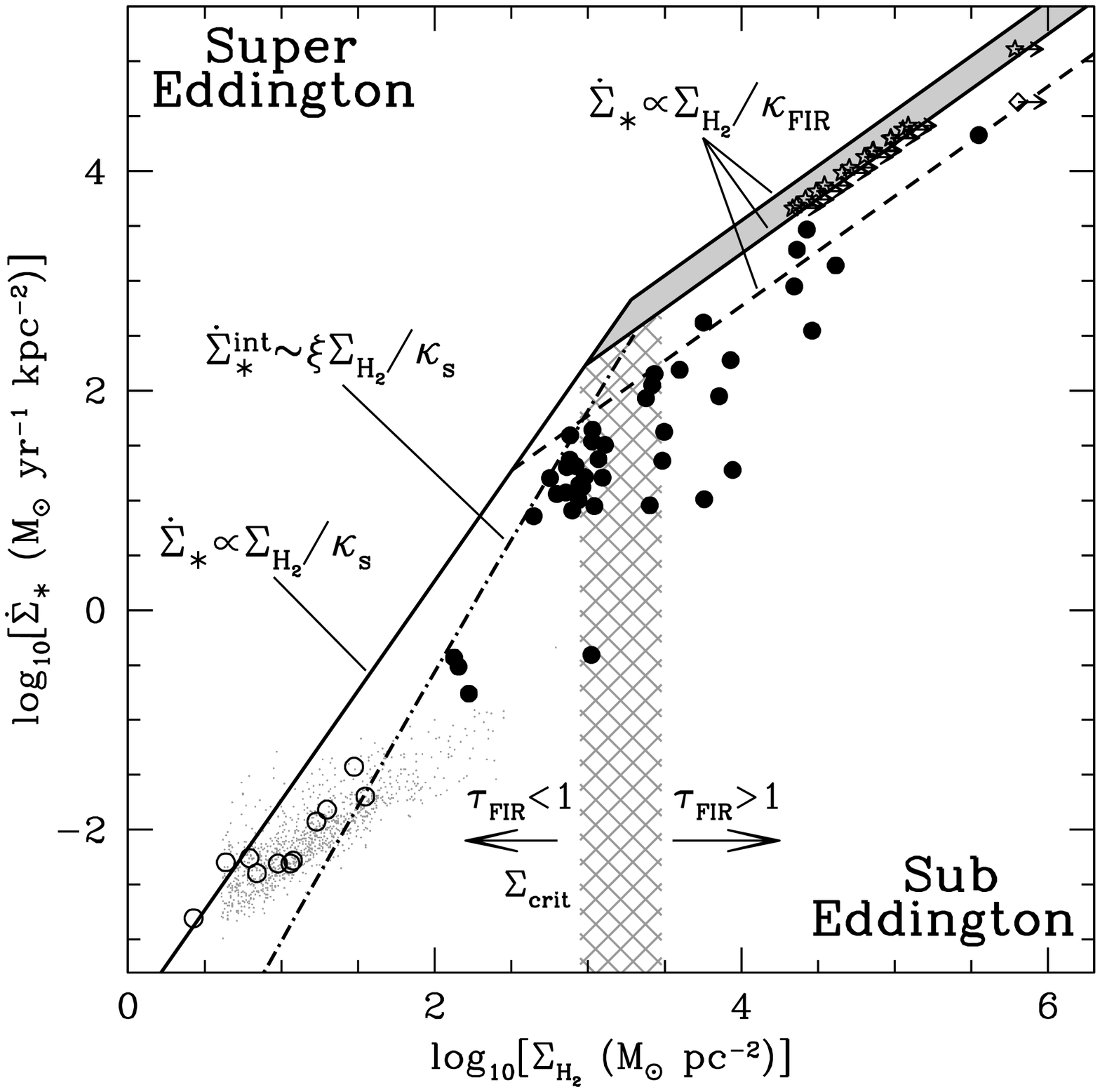,width=9cm}\psfig{file=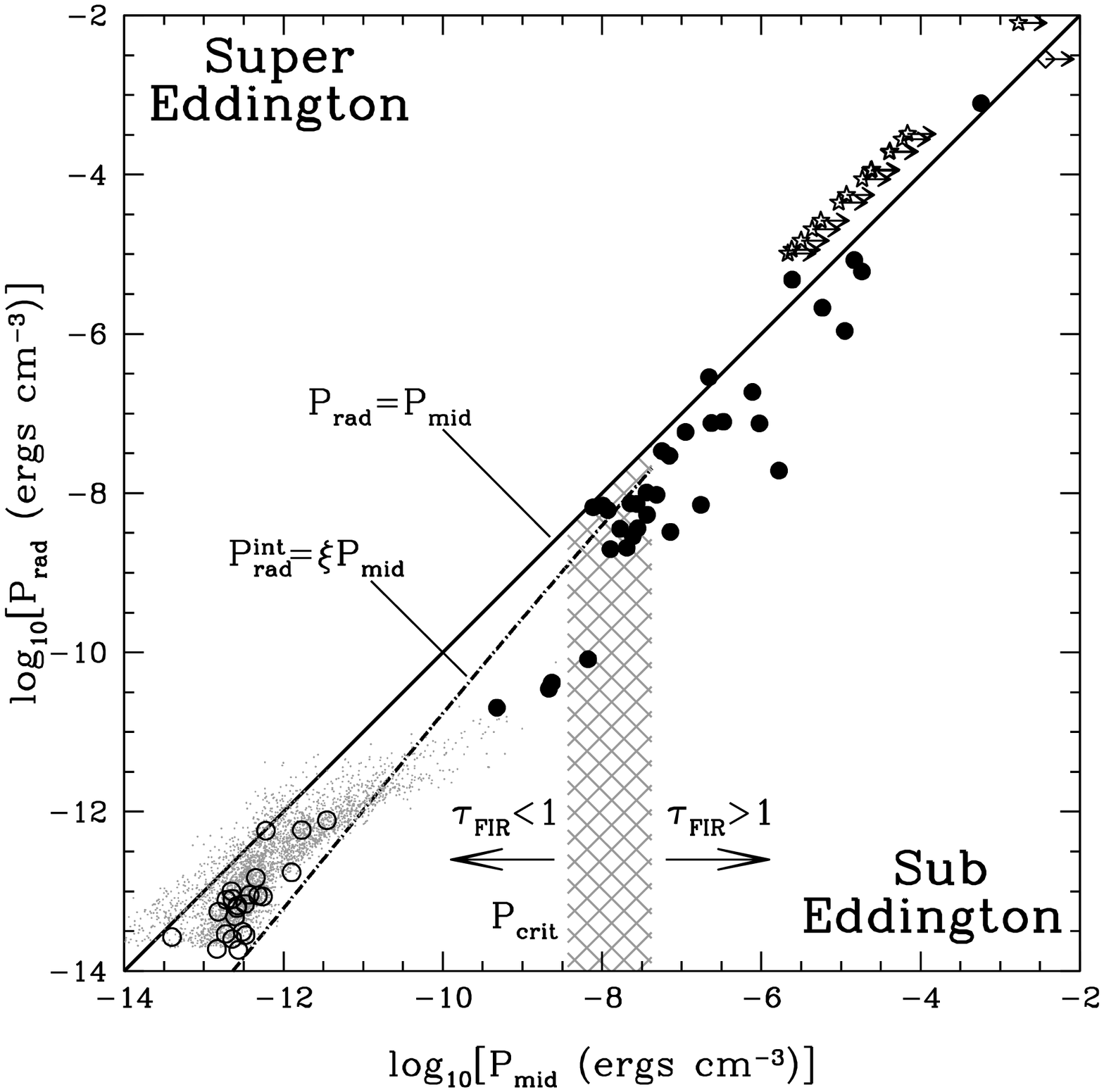,width=9cm}}
\caption{Star formation rate surface density ($\dot{\Sigma}_\star$) as a
function of the molecular gas surface density ($\Sigma_{\rm H_2}$) (left panel)
and radiation pressure as a function of midplane pressure ($P_{\rm mid}$;
eq.~\ref{eqn:pradpmid}) (right panel). The different symbols represent 750 pc
apertures of THINGS galaxies (small dots), THINGS galaxies (open circles),
starburst galaxies (solid circles), M82 super star clusters (stars), and the
Galactic Center star cluster (diamond).  The solid line in the
$\dot{\Sigma}_{\star}$--$\Sigma_{\rm H_2}$ plot is the Eddington limit for the
single-scattering ($\kappa=\kappa_{\rm s}$) limit.  The shaded region
corresponds to the optically thick Eddington limit for our preferred value of
the Rosseland-mean opacity ($\kappa_{\rm FIR}=5$--$10 \; {\rm cm^2 \; g^{-1}} \;
f_{\rm dg,150}$). The dashed line shows the effect of a factor of 3 increase in
the dust-to-gas ratio for the optically thick Eddington limit ($\kappa_{\rm
FIR}=30 \; {\rm cm^2 \; g^{-1}} \; f_{\rm dg,50}$).  In the $P_{\rm
rad}$--$P_{\rm mid}$ plot, the solid line shows the Eddington limit adopting
$\kappa_{\rm FIR}=10 \; {\rm cm^2\;g^{-1}} \; f_{\rm dg,150}$ for optically
thick gas.  The dot-dashed lines (both panels) are the intermittent Eddington
limit (eq.~\ref{eqn:int}). The hatched regions (both panels) are the critical
surface density or pressure for $h=30$--$100 \; \rm pc$ (eq.~\ref{eqn:sigcrit})
where $t_{\rm MS} \sim 2 t_{\rm dyn}$ and $\tau_{\rm FIR} \sim 1$. $\Sigma_{\rm
H_2}$ was calculated from $L_{\rm CO}^\prime$ using $X_{\rm CO}^{\rm MW}$ if
$L_{\rm IR} < 10^{11} \; \rm L_{\odot}$ and $X_{\rm CO}^{\rm ULIRG}$ if $L_{\rm
IR} > 10^{11} \; \rm L_{\odot}$. Overall, the Eddington limit suggests that
radiation pressure sets an upper bound to the $\dot{\Sigma}_{\star}$ or the
radiation pressure of a star-forming region or galaxy. Most star-forming regions
or galaxies are sub-Eddington, but a few THINGS apertures and optically thick
starbursts are super-Eddington (for $\kappa=10 \; {\rm cm^2\;g^{-1}} \; f_{\rm
dg,150}$). Several more optically thick starbursts will be consistent with or
even exceed the Eddington limit for $\kappa=30 \; {\rm cm^2\;g^{-1}} \; f_{\rm
dg,50}$. The rough agreement between starburst galaxies and the intermittent
Eddington limit reinforces the likely importance of intermittency. However, the
intermittent Eddington limit mildly under-predicts $\dot{\Sigma}_{\star}$ and
$P_{\rm rad}$ at for $\Sigma_{\rm H_2} \lesssim 10 \; \rm M_\odot pc^{-2}$ and
$P_{\rm mid} \lesssim 10^{-12} \; \rm ergs \; cm^{-3}$, indicating that the
effect of intermittency may be overestimated.
\label{fig:kslaw_pradpmid}}
\end{figure*}
%%%%%%%%%%%%%%%%%%%%%%%%%%%%%%%%%%%%%%%%%%%%%%%%%%%%%%%%%%%%%%%%%%%%%%%%%%%%
%%%%%%%%%%%%%%%%%%%%%%%%%%%%%%%%%%%%%%%%%%%%%%%%%%%%%%%%%%%%%%%%%%%%%%%%%%%%
\begin{figure*}
\centerline{\psfig{file=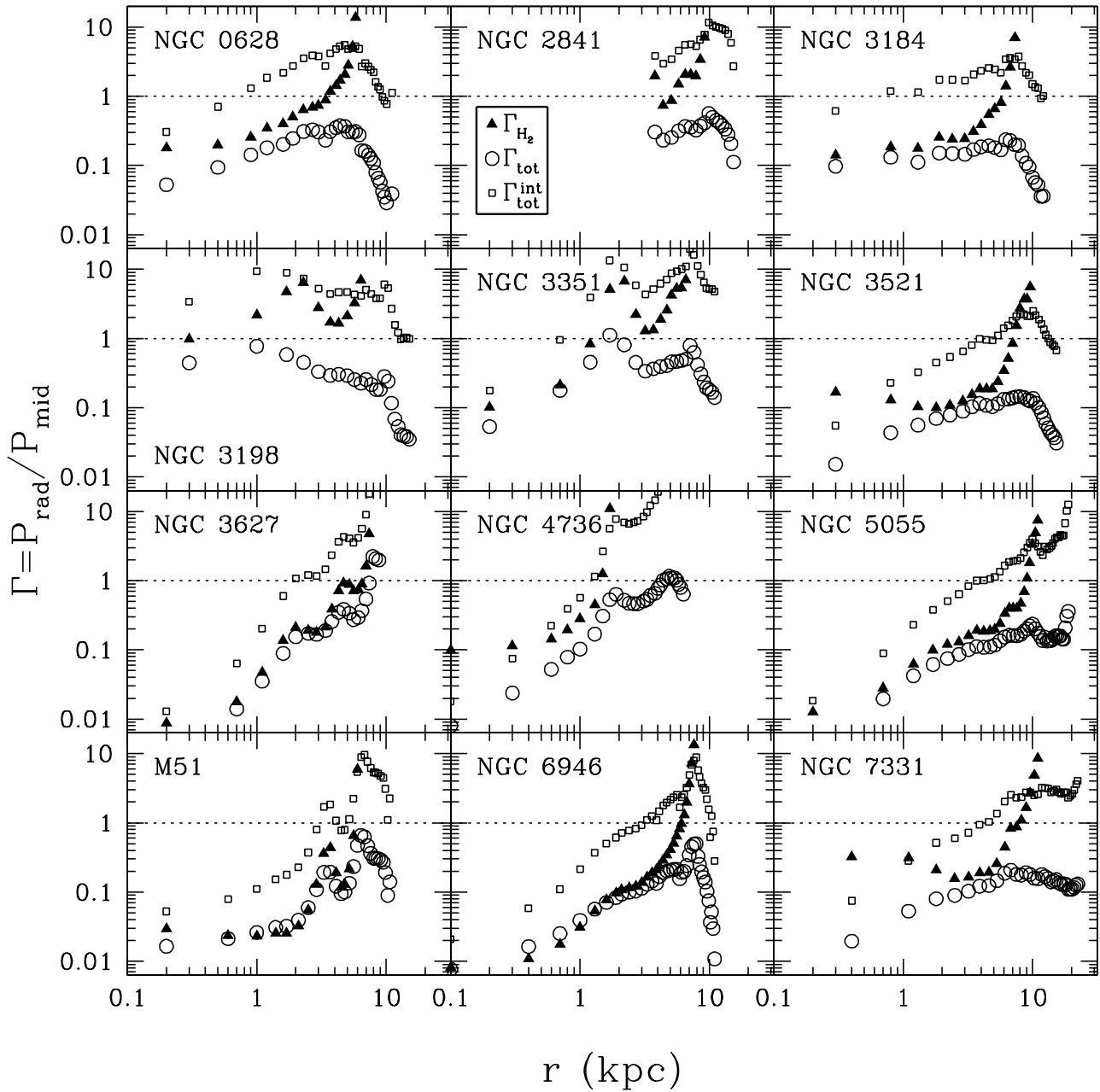,width=18cm}}
\caption{The Eddington ratio ($\Gamma=P_{\rm rad}/P_{\rm mid}$) as a function of
radius for THINGS galaxies with $\rm H_2$ detections. $\Gamma_{\rm H_2}$
($=P_{\rm rad}/(\pi G \Sigma_{\rm H_2}^2)$; solid triangles) and $\Gamma_{\rm
tot}$ ($=P_{\rm rad}/(0.5 \pi G \Sigma_{\rm g} \Sigma_{\rm tot})$; open circles)
represent two ways to calculate the midplane pressure. The open squares
($\Gamma_{\rm tot}^{\rm int}$) show the effect of adjusting $\Gamma_{\rm tot}$
for intermittency (see eq.~\ref{eqn:int}). $\Gamma_{\rm tot}$ tends to be
sub-Eddington, rising to a peak at $r \sim 5$--$10 \; \rm kpc$, and then rapidly
falling off. However, $\Gamma_{\rm tot}^{\rm int}$ is super-Eddington for $r
\gtrsim 1 \; \rm kpc$ in most of the galaxies, suggesting that $\xi$ likely
overestimates the importance of intermittency. $\Gamma_{\rm H_2}$ generally
follows the trend of $\Gamma_{\rm tot}$ in the inner regions of galaxies but
increases to super-Eddington values as $\Sigma_{\rm H_2}$ nears the detection
threshold.
\label{fig:radprofiles}}
\end{figure*}
%%%%%%%%%%%%%%%%%%%%%%%%%%%%%%%%%%%%%%%%%%%%%%%%%%%%%%%%%%%%%%%%%%%%%%%%%%%%
%%%%%%%%%%%%%%%%%%%%%%%%%%%%%%%%%%%%%%%%%%%%%%%%%%%%%%%%%%%%%%%%%%%%%%%%%%%%
\begin{figure}
\centerline{\psfig{file=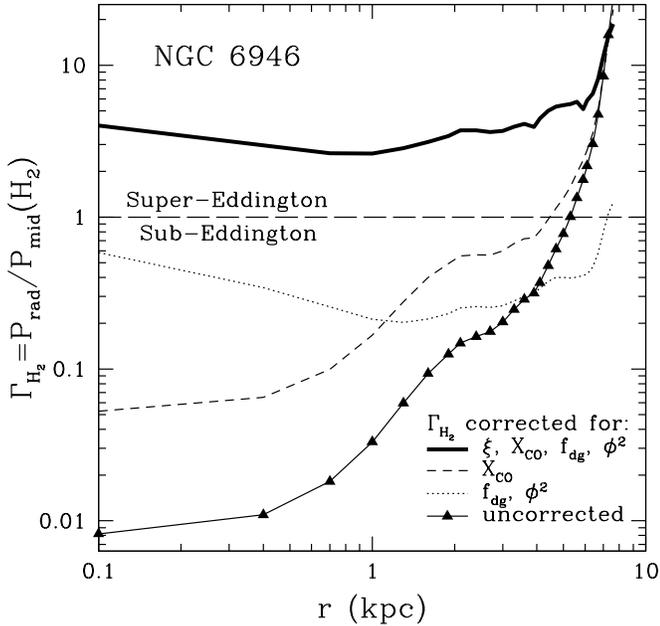,width=9cm}}
\caption{The molecular gas Eddington ratio $\Gamma_{\rm H_2}=P_{\rm rad}/(\pi G
 \Sigma_{\rm H_2}^2)$ as a function of radius for NGC 6946. The line styles show
 the uncorrected Eddington ratio (triangles and thin solid line) and
 $\Gamma_{\rm H_2}$ corrected for an $X_{\rm CO}$ gradient (dashed line), for a
 dust-to-gas ratio gradient plus a factor of $\phi^2$ ($\phi=h/R_{\rm GMC}$; see
 \S \ref{sec:intermittency}; dotted line), and for $X_{\rm CO}$ and dust-to-gas
 ratio gradients plus the intermittency factor and the $\phi^2$ factor (thick
 solid line). After all of these factors are accounted for, the Eddington ratio
 is $\sim$1 and nearly flat as a function of radius. We only show the effects of
 the $X_{\rm CO}$ and dust-to-gas ratio gradients on the Eddington ratio in NGC
 6946, but the profiles from the other THINGS galaxies shown in Figure
 \ref{fig:radprofiles} are qualitatively similar.
\label{fig:radprofiles6946}}
\end{figure}
%%%%%%%%%%%%%%%%%%%%%%%%%%%%%%%%%%%%%%%%%%%%%%%%%%%%%%%%%%%%%%%%%%%%%%%%%%%%
%%%%%%%%%%%%%%%%%%%%%%%%%%%%%%%%%%%%%%%%%%%%%%%%%%%%%%%%%%%%%%%%%%%%%%%%%%%%
\begin{figure}
\centerline{\psfig{file=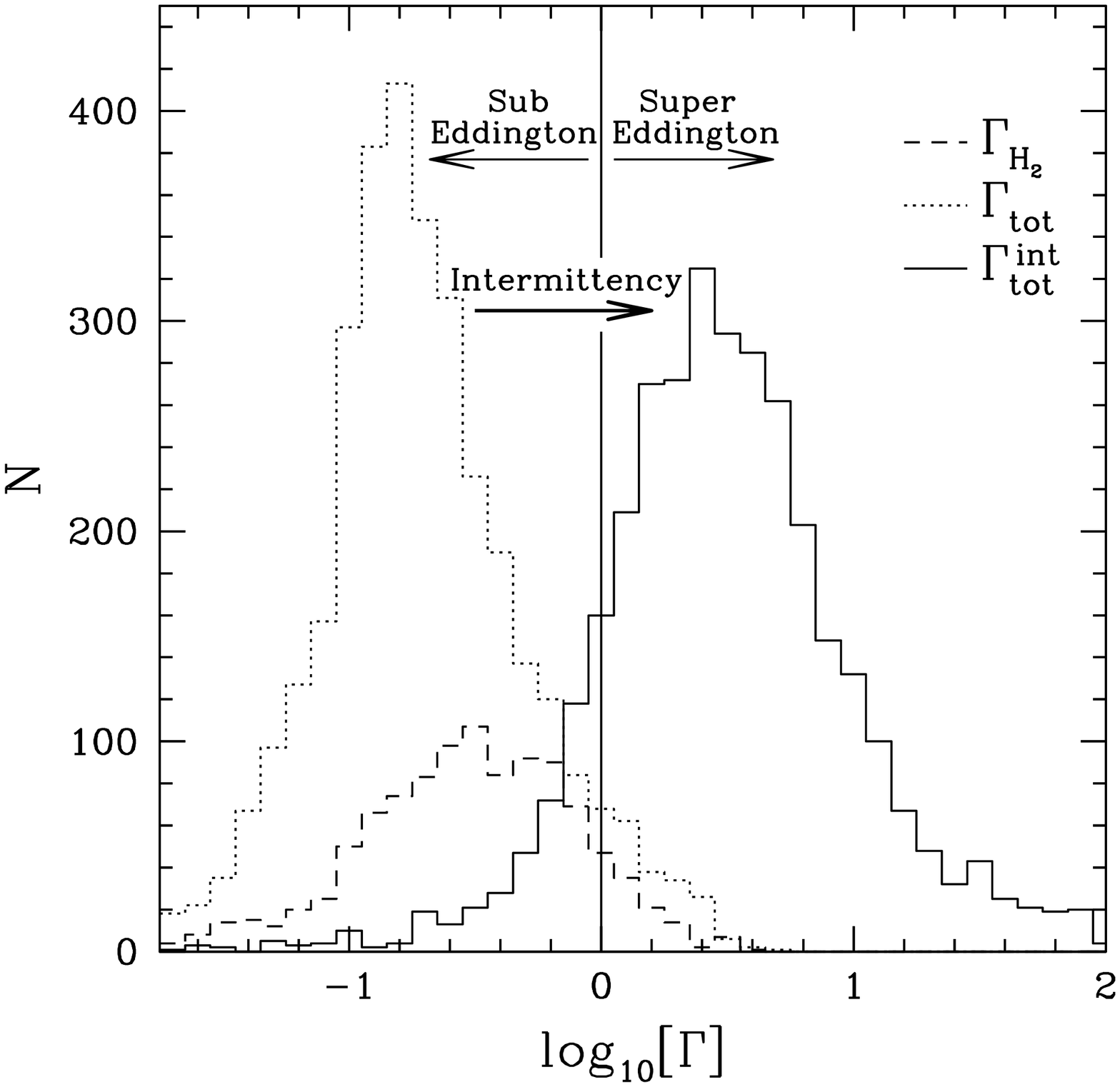,width=9cm}}
\caption{Histogram of the Eddington ratio ($\Gamma=P_{\rm rad}/P_{\rm mid}$) for
750 pc apertures of THINGS galaxies. The midplane pressure was calculated using
two different methods: $\Gamma_{\rm H_2}=P_{\rm rad}/(\pi G \Sigma_{\rm H_2}^2)$
(dashed line) and $\Gamma_{\rm tot}=P_{\rm rad}/(0.5 \pi G \Sigma_{\rm g}
\Sigma_{\rm tot})$ (dotted line). The solid line ($\Gamma_{\rm tot}^{\rm int}$)
shows the effect of adjusting the Eddington ratio by the intermittency factor
(eq.~\ref{eqn:int}). $\Gamma_{\rm tot}$ was calculated for all apertures with
either an H$_2$ or H\Rmnum{1} detection, but $\Gamma_{\rm H_2}$ could be
calculated only for apertures with H$_2$ detections. The $\Gamma_{\rm tot}$
distribution is mostly sub-Eddington, but intermittency pushes the majority of
the $\Gamma_{\rm tot}^{\rm int}$ distribution above the Eddington limit,
implying that $\xi$ may overestimate the importance of intermittency. The
$\Gamma_{\rm H_2}$ distribution is less peaked and shifted to higher Eddington
ratios than the $\Gamma_{\rm tot}$ distribution. Some apertures in the
$\Gamma_{\rm H_2}$ distribution are at or above the Eddington limit in spite of
the observations not being able to resolve individual star-forming regions.
\label{fig:inthist}}
\end{figure}
%%%%%%%%%%%%%%%%%%%%%%%%%%%%%%%%%%%%%%%%%%%%%%%%%%%%%%%%%%%%%%%%%%%%%%%%%%%%

We compile data of super star clusters, normal star-forming galaxies,
local starburst galaxies, ULIRGs, sub-millimeter galaxies (SMGs), hyper luminous
infrared galaxies, and circumnuclear starbursts to assess feedback from
radiation pressure. Below, we test the hypothesis that radiation pressure is
dynamically important in galaxies and star-forming subregions by comparing data
to the Eddington limit (\S \ref{sec:theory}) on a variety of physical scales
ranging from globally-averaged properties of galaxies to individual star-forming
subregions within galaxies.

\subsection{IR Luminosity vs.~Molecular Line Luminosity}
\label{sec:lirlcolhcn}
We show the total IR luminosity $L_{\rm IR}$ as a function of molecular line
luminosity\footnote{We used the {\it J} = 1--0 line unless only higher order
lines were available.} $L_{\rm CO}^\prime$ (Figure \ref{fig:lirlco}) and $L_{\rm
HCN}^\prime$ (Figure \ref{fig:lirlhcn}) for our sample of star-forming
galaxies.\footnote{Aravena et al.~2008; Becklin et al.~1980; Beelen et al.~2006;
Benford et al.~1999; Capak et al.~2008; Carilli et al.~2005; Casoli et al.~1989;
Chapman et al.~2005; Chung et al.~2009; Combes et al.~2010; Coppin et al.~2009;
Coppin et al.~2010; Daddi et al.~2007; Daddi et al.~2009; Daddi et al.~2010a;
Downes \& Solomon 1998; Gao et al.~2007; Gao \& Solomon 1999; Gao \& Solomon
2004a,b; Genzel et al.~2003; Graci{\'a}-Carpio et al.~2008; Greve et al.~2005;
Greve et al.~2006; Isaak et al.~2004; Kim \& Sanders 1998; Knudsen et al.~2007;
Mauersberger et al.~1996; Mirabel et al.~1990; Momjian et al.~2007; Murphy et
al.~2001; Neri et al.~2003; Riechers et al.~2006; Riechers et al.~2007; Riechers
et al.~2008; Sajina et al.~2008; Sakamoto et al.~2008; Sanders et al.~1991;
Schinnerer et al.~2006; Schinnerer et al.~2007; Schinnerer et al.~2008; Smith \&
Harvey 1996; Solomon et al.~1997; Solomon \& Vanden Bout 2005; Walter et
al.~2003; Walter et al.~2009; Wei{\ss} et al.~2001; Yan et al.~2010; Young \&
Scoville 1982; Yun et al.~2001.} $L_{\rm IR}$ is known to trace the total light
from massive stars (e.g., Kennicutt 1998),\footnote{In reality, for normal
galaxies a fraction of the UV and optical light escapes before being reprocessed
by dust, and a fraction of the IR is diffuse and likely not associated with star
formation (e.g., Kennicutt et al.~2010, Calzetti et al.~2010).  The UV and IR
luminosities are roughly equal at a bolometric luminosity of $L_{\rm bol} \sim
10^{10} \, \rm L_\odot$, but the UV luminosity is an order of magnitude larger
than the IR luminosity at $L_{\rm bol} \sim 10^{8.5} \, \rm L_\odot$ (Martin et
al.~2005). Thus, we expect that galaxies with $L_{\rm IR} \lesssim 10^{10} \,
\rm L_\odot$ to move closer to the Eddington limit in Figure \ref{fig:lirlco}.}
whereas $L_{\rm CO}^\prime$ and $L_{\rm HCN}^\prime$ provide a measure of the
total gas mass and dense gas mass, respectively. Under the assumption that the
total gravitational potential is dominated by the gas on the physical scales
where the stars are forming, the Eddington luminosity is related to $L_{\rm
CO}^\prime$ by
\begin{equation}
 L_{\rm Edd} = \frac{4\pi G c}{\kappa} X_{\rm CO} \, L_{\rm CO}^\prime,
\label{eqn:leddlco}
\end{equation}
where $X_{\rm CO}$ is the $L_{\rm CO}^\prime$-to-$M_{\rm H_2}$ conversion factor
and $\kappa$ is the appropriate flux-mean or Rosseland-mean opacity (either
single-scattering or optically thick; see
\S\ref{sec:sslimit}--\ref{sec:optthick}).  Although counterintuitive, the
single-scattering Eddington luminosity lies below the optically thick Eddington
limit because the dust opacity is column-averaged and highly non-grey (see \S
\ref{sec:radpressure}).  We adopt $X_{\rm CO}^{\rm MW} = 4.4 \; \rm
M_{\odot}(K\;km\;s^{-1} \;pc^2)^{-1}$ for normal galaxies (including a
correction factor of 1.36 to account for He; Strong \& Mattox 1996; Dame et
al.~2001), and $X_{\rm CO}^{\rm ULIRG} = 0.8 \; \rm M_{\odot} (K\;km\;s^{-1}
\;pc^2)^{-1}$ for galaxies with $L_{\rm IR} \geq 10^{11} L_{\odot}$ (as
appropriate for starbursts and ULIRGs; e.g., Downes \& Solomon 1998).
Similarly, if the majority of the IR luminosity comes from regions where the
dense molecular gas dominates the potential, then $L_{\rm Edd}$ is related to
$L_{\rm HCN}^\prime$ by
\begin{equation}
 L_{\rm Edd} = \frac{4\pi G c}{\kappa_{\rm FIR}} X_{\rm HCN} \, L_{\rm
 HCN}^\prime,
\label{eqn:leddlhcn}
\end{equation}
where we explicitly write $\kappa=\kappa_{\rm FIR}$ because the critical density
for HCN emitting gas is large enough that these regions should always be
optically-thick (\S \ref{sec:optthick}).\footnote{Using $\rho_{\rm crit, \, HCN}
\sim 10^{-19} \; \rm g \; cm^{-3}$, $\tau_{\rm FIR}\sim\kappa_{\rm FIR}\rho_{\rm
crit, \, HCN}R$ is larger than unity for scales $R\gtrsim$\,pc and $\kappa_{\rm
FIR}\gtrsim$\,few cm$^2$ g$^{-1}$.}  In eq.~\ref{eqn:leddlhcn}, we take an
$L_{\rm HCN}^\prime$-to-$M_{\rm H_2}^{\rm dense}$ conversion factor of $X_{\rm
HCN}=3 \;\rm M_{\odot} (K \;km \;s^{-1} \;pc^2)^{-1}$, but we caution that
$X_{\rm HCN}$ is uncertain to a factor of $\sim$3 (Gao \& Solomon 2004a,b; see
\S \ref{sec:xcoxhcn}).

In Figure \ref{fig:lirlco}, the lines indicate the Eddington limit for various
limiting cases.  The lines in the left panel show the single-scattering
Eddington limit (solid line) and the single-scattering Eddington limit
accounting for intermittency (dot-dashed line) assuming $h = 100 \, \rm pc$ and
$r = 10 \, \rm kpc$ (eq.~\ref{eqn:leddlco}).  The shaded region in the right
panel is the optically thick Eddington limit (eq.~\ref{eqn:leddlco} with
$\kappa_{\rm FIR} = 5$--$10 \; \rm cm^2 \; g^{-1}$) and the dashed line shows
the optically thick Eddington limit for an enhanced opacity ($\kappa_{\rm FIR} =
30 {\rm \; cm^{2} \; g^{-1}} \; f_{\rm dg,50}$; where $f_{\rm dg, 50} = f_{\rm
dg} \times 50$) due to an assumed higher dust-to-gas ratio in dense star-forming
environments.  We plot the single-scattering (left panel) and optically thick
(right panel) Eddington limits separately for clarity, but the data are the same
in both panels.  The different symbols indicate various rotational transitions
of CO: solid circles ({\it J} = 1--0), crosses ({\it J} = 2--1), squares ({\it
J} = 3--2), and triangles ({\it J} = 4--3).

Note that no galaxies exceed the optically thick Eddington limit and most
galaxies are neither significantly super- or sub-Eddington with respect to the
single-scattering Eddington limit.  We caution that the applicable Eddington
limit for any individual galaxy cannot be determined in this plot due to the
lack of surface density measurements, which dictate the optical depth to the FIR
and the relevant Eddington limit.  For example, high surface density
star-forming regions can be optically thick at $L_{\rm CO}^\prime \lesssim 10^9
\rm \; K \; km \; s^{-1} \; pc^{2}$ and lie to the left of the single-scattering
Eddington limit (solid line in left panel) but below the optically thick
Eddington limit (shaded region in right panel).  For the single-scattering
limit, our assumption of $r=10$ kpc is accurate to a factor of $\sim$5 for most
of the sample, but the single-scattering opacity scales as $r^{-2}$, so it is
only accurate to a factor of $\sim$25. Some compact starbursts have radii much
smaller than our assumed radius, so they are optically thick, even at low
$L_{\rm CO}^\prime$ and this explains why they exceed the single-scattering
limit in the left panel but are below the optically thick limit in the the right
panel.  In addition, note that the optically-thick Eddington limit is a hard
upper bound to a galaxy's IR luminosity, which suggests that radiation pressure
feedback may set the maximum SFR of a galaxy. In the Eddington-limited model,
the scatter in the $L_{\rm IR}$--$L_{\rm CO}^\prime$ relation may be due to
variations in $h$, $X_{\rm CO}$ (see \S \ref{sec:xcoxhcn}), the dust-to-gas
ratio/metallicity (see \S \ref{sec:DGR}), the effective radii, and the depth of
the stellar potential.\footnote{Note that previous work by Krumholz \& Thompson
(2007) and Narayanan et al.~(2008) explains the slopes of the $L_{\rm
IR}$--$L_{\rm CO}^\prime$ and $L_{\rm IR}$--$L_{\rm HCN}^\prime$ relations by
comparing the critical density of the gas tracer to the median density of the
ISM.}

The intermittency of star formation will likely affect the Eddington limit for
CO-emitting gas (dot-dashed line in left panel). $L_{\rm CO}^\prime$ traces the
total molecular gas reservoir including the molecular gas that is not actively
participating in star formation, such as GMC envelopes and diffuse intercloud
gas.  The gas mass relevant for the Eddington limit may be overestimated for
galaxies in the single-scattering limit. To account for this, we multiply the
Eddington luminosity by the intermittency factor for CO-emitting gas $\xi \sim
0.06$ for the Milky Way value of $X_{\rm CO}$, $h=100\; \rm pc$, $r=10 \; \rm
kpc$, and $L_{\rm CO}^\prime=10^9 \rm \; K\;km\;s^{-1} \;pc^2$ (see
eq.~\ref{eqn:int} and \S \ref{sec:intermittency}).  The intermittency factor
approaches unity when $L_{\rm CO}^\prime \sim 2 \times 10^{11} {\rm \;
K\;km\;s^{-1} \;pc^2} \; h_{100} \, r_{10}^2/X_{\rm CO}^{\rm MW}$.  Thus,
compact star-forming regions, such as the nuclear starbursts of ULIRGs, have
$\xi\sim1$ at low $L_{\rm CO}^\prime$ due to their very small radii (e.g.,
Downes \& Solomon 1998).

Figure \ref{fig:lirlhcn} shows the $L_{\rm IR}$--$L_{\rm HCN}^\prime$ relation
for our sample of star-forming galaxies.  The shaded region represents the
optically-thick Eddington limit (eq.~\ref{eqn:leddlhcn} with $\kappa_{\rm FIR} =
5$--$10 \; \rm cm^2 \; g^{-1}$) and the dashed line shows the optically thick
Eddington limit for an enhanced opacity ($\kappa_{\rm FIR} = 30 {\rm \; cm^{2}
\; g^{-1}} \; f_{\rm dg,50}$; where $f_{\rm dg, 50} = f_{\rm dg} \times 50$),
which may result from a higher dust-to-gas ratio in dense star-forming
environments.  The circles and crosses correspond to the {\it J} = 1--0 and the
{\it J} = 2--1 rotational transitions of HCN, respectively.

The $L_{\rm IR}$--$L_{\rm HCN}^\prime$ relation (Figure \ref{fig:lirlhcn}) is
tight and linear over several orders of magnitude, implying that stars form out
of dense gas (Gao \& Solomon 2004a,b; Wu et al.~2005). The dense gas fraction
($L_{\rm HCN}^\prime$/$L_{\rm CO}^\prime$) is nearly constant for galaxies with
$L_{\rm IR} \lesssim 10^{11} L_{\odot}$ (Gao \& Solomon 2004b), so $L_{\rm
CO}^\prime$ can be used to indirectly trace the dense gas mass $M_{\rm H_2}^{\rm
dense}$. However, the dense gas fraction increases dramatically in LIRGs and
ULIRGs ($L_{\rm IR} \gtrsim 10^{11} L_{\odot}$), so CO does not trace dense gas
mass in these galaxies (Gao \& Solomon 2004b). HCN, on the other hand, has a
critical density for excitation that is $\sim$2 orders of magnitude larger than
that of CO, so it traces dense, optically thick gas in star-forming GMC cores
rather than diffuse GMC envelopes. The dynamical time for HCN-emitting gas is
much less than the main-sequence lifetime of the most massive stars, so the
intermittency factor for HCN-emitting gas will be approximately unity:
\begin{equation}
 t_{\rm dyn}^{\rm HCN} \sim 2 \times 10^5 \; {\rm yr} \; \rho_{\rm crit, \,
 HCN}^{-1/2} \ll t_{\rm MS} \rightarrow \xi \approx 1,
\end{equation}
where $\rho_{\rm crit, \, HCN} \sim 10^{-19} \; \rm g \; cm^{-3}$.

If the picture of radiation pressure feedback is correct, then it should
determine the $L_{\rm IR}$--$L_{\rm HCN}^\prime$ correlation directly.  In fact,
both the Eddington limit and the data show a linear relation between $L_{\rm
IR}$ and $L_{\rm HCN}^\prime$.  The galaxies closely follow but do not exceed
the Eddington limit for our preferred value of the Rosseland-mean opacity
($\kappa_{\rm FIR}=5$--$10 {\rm \; cm^{2} \; g^{-1}} \; f_{\rm dg,150}$). If the
opacity is higher ($\kappa_{\rm FIR}=30 {\rm \; cm^{2} \; g^{-1}} \; f_{\rm dg,
50}$), then many galaxies are consistent with Eddington and a number are
super-Eddington. For any of the values of the opacity that we assume, the
general agreement between $L_{\rm IR}$ and $L_{\rm HCN}^\prime$ suggests that
radiation pressure may play an important role in regulating star formation
(Scoville et al. 2003). However, a number of important factors remain uncertain,
which we discuss in \S \ref{sec:discussion}.

\subsection{Molecular Schmidt Law and Radiation Pressure}
\label{sec:kslaw-prad}
The Schmidt law is a tight power law relation between the surface density of
star formation rate $\dot{\Sigma}_{\star}$ and the gas surface density
($\dot{\Sigma}_{\star} \propto \Sigma_{\rm g}^{1.4}$; Kennicutt
1998). Furthermore, Bigiel et al.~(2008) found that the Schmidt law for
molecular gas is linear within local star-forming galaxies
($\dot{\Sigma}_{\star} \propto \Sigma_{\rm H_2}^{1.0}$). In the left panel of
Figure \ref{fig:kslaw_pradpmid}, we plot $\dot{\Sigma}_{\star}$ vs.~$\Sigma_{\rm
H_2}$ for individual apertures of THINGS galaxies (small dots; Bigiel et
al.~2008; Leroy et al.~2008), THINGS galaxies with $\rm H_2$ detections (open
circles), starburst galaxies \footnote{Aravena et al.~2008; Becklin et al.~1980;
Benford et al.~1999; Capak et al.~2008; Casoli et al.~1989; Chapman et al.~2005;
Coppin et al.~2009; Coppin et al.~2010; Daddi et al.~2009; Downes \& Eckart
2007; Downes \& Solomon 1998; Greve et al.~2005; Knudsen et al.~2007;
Mauersberger et al.~1996; Momjian et al.~2007; Neri et al.~2003; Paumard et
al.~2006; Riechers et al.~2007; Riechers et al.~2008; Sajina et al.~2008;
Sakamoto et al.~2008; Schinnerer et al.~2006; Schinnerer et al.~2007; Schinnerer
et al.~2008; Smith \& Harvey 1996; Walter et al.~2003; Walter et al.~2009;
Wei{\ss} et al.~2001; Yan et al.~2010; Young \& Scoville 1982; Yun et al.~2001}
(solid circles), M82 super star clusters (stars; McCrady et al.~2003; McCrady \&
Graham 2007), and the Galactic Center star cluster (diamond; Paumard et
al.~2006). We compare the data with the Eddington limit using
$\dot{\Sigma}_{\star}$ and $\Sigma_{\rm H_2}$ as proxies for the radiation and
gravitational forces,
\begin{equation}
 \dot{\Sigma}_{\star}^{\rm Edd} = \frac{4 \pi G \Sigma_{\rm H_2}}{\epsilon c
 \kappa},
\label{eqn:schmidtlaw}
\end{equation}
where $\epsilon \approx 5 \times 10^{-4}$ is the efficiency of the conversion of
mass into luminosity during the star formation process assuming a Kroupa (2001)
broken power law IMF that extends up to $120 \; \rm M_\odot$. For star clusters
we assume a light-to-mass ratio appropriate for a zero age main sequence stellar
population ($\Psi=3000 \; \rm ergs \; s^{-1} g^{-1}$) and that the stellar mass
is a lower limit on the gas mass of the parent GMC.  We use the same $X_{\rm
CO}$ values as in Figure \ref{fig:lirlco} and again caution
that $X_{\rm CO}$ is uncertain to a factor of a few and may vary systematically
from normal galaxies to starbursts (see \S \ref{sec:xcoxhcn}).

For the molecular Schmidt law, we find that the Eddington limit is an upper
bound to $\dot{\Sigma}_{\star}$.  Most star-forming regions and galaxies follow
the Eddington limit (solid line) and are within $\sim$1.5 dex of it.  The
Eddington limit accounting for intermittency (dot-dashed line) appears to agree
better with the data than the naive single-scattering Eddington limit,
suggesting that intermittency may be an important effect.  As the medium becomes
optically thick near the critical surface density for intermittency (hatched
region; see \S \ref{sec:intermittency}), the optically thick Eddington limit
($\dot{\Sigma}_{\star}^{\rm Edd} \propto \Sigma_{\rm H_2}/\kappa_{\rm FIR}
\propto \Sigma_{\rm H_2}^{1.0}$) provides a firm upper bound to
$\dot{\Sigma}_{\star}$ for our preferred value of the dust opacity ($\kappa_{\rm
FIR} = 5$--$10 \; {\rm cm^2 \; g^{-1}} \; f_{\rm dg,150}$). If the dust opacity
($\kappa_{\rm FIR} = 30 \; {\rm cm^2 \; g^{-1}} \; f_{\rm dg,50}$) is enhanced
due to a higher assumed dust-to-gas ratio, then some galaxies reach the
optically thick Eddington limit and a few galaxies exceed it.

In the right panel of Figure \ref{fig:kslaw_pradpmid}, we plot the radiation
pressure from UV and FIR photons versus the midplane pressure. These pressures
will balance each other at Eddington (solid line):
\begin{equation}
 P_{\rm rad} \sim (1+\tau_{\rm FIR})\frac{F}{c} \sim P_{\rm mid} = \frac{\pi}{2}
 G \Sigma_{\rm g} \Sigma_{\rm tot},
 \label{eqn:pradpmid}
\end{equation}
where we take $\Sigma_{\rm tot} \equiv \Sigma_{\rm g}+0.1\Sigma_{\star}$ (Wong
\& Blitz 2002). The $P_{\rm rad}$--$P_{\rm mid}$ plot shows that radiation
pressure correlates strongly with midplane pressure over 10 orders of
magnitude. The Eddington limit serves as a rough upper limit to $P_{\rm rad}$,
and most galaxies are within 2 dex of the Eddington limit. We note that some of
the THINGS apertures and some dense starbursts reach or exceed the Eddington
limit. For galaxies with $P_{\rm mid} < P_{\rm crit}$, the critical midplane
pressure (hatched region; see \S \ref{sec:intermittency} \&
\ref{sec:inter-disc}), we expect that the effects of intermittency are
important; however, the intermittency adjusted Eddington limit (dot-dashed line)
under-predicts $P_{\rm rad}$ for star-forming regions with $P_{\rm mid} \lesssim
10^{-11.5} \rm \; ergs \; cm^{-3}$. The intermittency factor may overestimate
the importance of intermittency because of the simplifying assumption that
subregions are ``on'' or ``off'' (see \S \ref{sec:intermittency}). We also see
that galaxies and star-forming regions with $10^{-11} \; {\rm ergs \; cm^{-3}}
\lesssim P_{\rm mid} \lesssim P_{\rm crit}$ tend to fall significantly below the
Eddington limit, possibly because our simple parametrization of $X_{\rm CO}$
(see \S \ref{sec:xcoxhcn}) is overestimating $M_{\rm H_2}$ (and $P_{\rm mid}$)
for these systems. Radiation pressure becomes increasingly more important in the
optically thick limit ($P_{\rm rad} \gtrsim P_{\rm crit}$) as some galaxies and
star-forming regions meet and exceed Eddington. As expected from Figures
\ref{fig:lirlco} \& \ref{fig:lirlhcn}, if we assume a larger dust-to-gas ratio
and opacity ($\kappa=30 \; {\rm cm^2 \; g^{-1}} \; f_{\rm dg, 50}$) potentially
appropriate for dusty galaxies, then more of the optically thick starbursts
would be super-Eddington.

\subsection{Radiation Pressure on Sub-galactic Scales}
\label{sec:subgalactic_scales}
So far we have evaluated radiation pressure on a galaxy-wide scale; however, the
distribution of gas and star formation in galaxies is
inhomogeneous. Consequently, the Eddington ratio ($\Gamma=P_{\rm rad}/P_{\rm
mid}$) will likely vary on sub-galactic scales. We use observations from the
THINGS survey (Walter et al.~2008; Leroy et al.~2008; Bigiel et al.~2008) to
calculate the Eddington ratio as a function of radius in azimuthally averaged
radial bins and for semi-resolved (750 pc) apertures. Since the THINGS galaxies
are generally in the single-scattering limit (see eq.~\ref{eqn:sslimit}), we
conservatively adopt the radiation pressure to be $P_{\rm rad}^{\rm IR}=F_{\rm
IR}/c$ (see eq.~\ref{eqn:pradpmid}). For the midplane pressure given in
eq.~\ref{eqn:pradpmid}, the corresponding Eddington ratio is $\Gamma_{\rm
tot}=P_{\rm rad}^{\rm IR}/(0.5 \pi G \Sigma_{\rm g} \Sigma_{\rm tot})$. Stars
and atomic gas may not contribute significantly to the surface density in
regions of active star formation, so we also calculate the Eddington ratio
assuming that the midplane pressure depends only on the total gas surface
density $\Gamma_{\rm g}=P_{\rm rad}^{\rm IR}/(\pi G \Sigma_{\rm g}^2)$ or the
molecular gas surface density $\Gamma_{\rm H_2}=P_{\rm rad}^{\rm IR}/(\pi G
\Sigma_{\rm H_2}^2)$. Intermittency may be important because the THINGS
observations cannot resolve individual star-forming regions. We calculate the
Eddington ratio corrected for intermittency ($\Gamma_{\rm tot}^{\rm
int}=\Gamma_{\rm tot}/\xi$, see eq.~\ref{eqn:int}). In Figure
\ref{fig:radprofiles}, we plot $\Gamma_{\rm tot}$ (open circles), $\Gamma_{\rm
H_2}$ (solid triangles), and $\Gamma_{\rm tot}^{\rm int}$ (open squares) as a
function of radius for azimuthally averaged rings. We find that $\Gamma_{\rm g}$
is similar to $\Gamma_{\rm tot}$, so we omit $\Gamma_{\rm g}$ for clarity.

At intermediate radii ($r \sim 1 \rightarrow \rm several \; kpc$), $\Gamma_{\rm
tot}$ and $\Gamma_{\rm H_2}$ generally increase from sub-Eddington ($\Gamma \sim
0.1$) to approaching or exceeding the Eddington limit ($\Gamma \sim
1$). $\Gamma_{\rm tot}$ reaches a maximum Eddington ratio at $r \sim 5$--$10 \;
\rm kpc$, where it falls off steeply. As the observations near the $\rm H_2$
detection threshold, $\Gamma_{\rm H_2}$ increases rapidly due to a small
$\Sigma_{\rm H_2}$ with large error bars (see, e.g., Figure 40 from Leroy et
al.~2008), and thus the large value of $\Gamma_{\rm H_2}$ at large $r$ is
consistent with Eddington to within the errors on $\Sigma_{\rm H_2}$.  For $r >
1 \rm \; kpc$ where $\Gamma_{\rm tot} < 1$, the intermittency factor can boost
$\Gamma_{\rm tot}^{\rm int}$ to the Eddington limit, suggesting that
intermittency is important.

The $\Gamma<1$ regions in the inner parts of star-forming galaxies present a
challenge for radiation pressure regulated star formation. Intermittency cannot
account for the low Eddington ratios of these regions.  However, a metallicity
gradient, as seen in observations of star-forming galaxies (Mu{\~n}oz-Mateos et
al.~2009), increases the Eddington ratio at small radii (see \S \ref{sec:DGR}).
A metallicity gradient that rises at smaller radii correlates with a decreasing
gradient in $X_{\rm CO}$ (Sodroski et al.~1995; Arimoto et al.~1996) and an
increasing gradient in the dust-to-gas ratio (Mu{\~n}oz-Mateos et al.~2009).  We
adopt the $X_{\rm CO}$ gradient given by eq.~10 of Arimoto et al.~(1996) for
data from the Milky Way, M31, and M51 (${\rm log}\,X/X_{\rm e}=0.41[r/r_{\rm
e}-1]$, where $r_{\rm e}$ is the effective radius, which we assume to be 7 kpc
and $X_{\rm e}$ is the value of $X_{\rm CO}$ at the effective radius).  To
account for the dust-to-gas ratio gradient, we use a power law interpolation
between $f_{\rm dg}=1/30$ at 0.1 kpc and $f_{\rm dg}=1/150$ at 10 kpc, motivated
by Figure 15 of Mu{\~n}oz-Mateos et al.~(2009).  In addition, collapsing GMCs
enhance the surface density by a factor of $\phi^2$ ($\phi=h/R_{\rm GMC}$; see
\ref{sec:intermittency}), making some regions optically thick to FIR radiation.
In Figure \ref{fig:radprofiles6946}, we show the molecular Eddington ratio as a
function of radius for NGC 6946 since this galaxy is well below ($\sim$2 dex)
the Eddington limit at small radii (see Figure \ref{fig:radprofiles}).  After
accounting for intermittency, an $X_{\rm CO}$ gradient, a dust-to-gas ratio
gradient, and a surface density enhancement in the GMCs, we find that
$\Gamma_{\rm H_2} \sim 1$ for almost all radii in NGC 6946.  We find
qualitatively similar results for all of the THINGS galaxies shown in Figure
\ref{fig:radprofiles} assuming that the metallicity, $X_{\rm CO}$, and
dust-to-gas ratio gradients are similar to those adopted for NGC 6946.  Thus, it
is at least in principle possible to explain the nominally sub-Eddington inner
regions of local star-forming galaxies using a combination of these effects.

In Figure \ref{fig:inthist}, we plot the distributions of the individual
Eddington ratios for the THINGS apertures (750 pc resolution): $\Gamma_{\rm
tot}$ (dotted line), $\Gamma_{\rm H_2}$ (dashed line), and $\Gamma_{\rm
tot}^{\rm int}$ (solid line). The distribution of $\Gamma_{\rm tot}$ is peaked
around $\Gamma_{\rm tot} \sim 0.1$, and the majority of apertures are
sub-Eddington for $\Gamma_{\rm tot}$. The high $\Gamma_{\rm tot}$ tail of the
distribution extends above the Eddington limit, with super-Eddington apertures
comprising 5\% of the total apertures and containing 5\% of the total
flux. Star-forming regions are unresolved on 750 pc scales, so $\Gamma_{\rm
tot}$ should be adjusted to account for intermittency ($\Gamma_{\rm tot}^{\rm
int}$). However, most apertures lie above the intermittency adjusted Eddington
limit. As in Figures \ref{fig:lirlco}--\ref{fig:radprofiles6946}, this shift
suggests that intermittency is important for radiation pressure supported
star formation in normal spirals; however, $\xi$ appears to overestimate the
importance of intermittency, possibly due to the simplifying assumption that
subregions are either ``on'' or ``off'' (see \S \ref{sec:intermittency} \&
\ref{sec:inter-disc}). The distributions of $\Gamma_{\rm tot}$ and $\Gamma_{\rm
g}$ are similar, so we do not plot $\Gamma_{\rm g}$ for clarity.

The distribution of $\Gamma_{\rm H_2}$ is less peaked and shifted to
systematically higher values than the distribution of $\Gamma_{\rm tot}$ with
10\% of these apertures radiating at or above the Eddington limit. This is not
surprising (given Figure \ref{fig:radprofiles}) because radiation pressure will
likely be more important in $\rm H_2$-dominated star-forming regions. The
detection limit for $\rm H_2$ is higher than that for H\Rmnum{1}, so the
distribution of $\Gamma_{\rm H_2}$ contains fewer apertures than $\Gamma_{\rm
tot}$. In addition, the apertures with $\rm H_2$ detections tend to be within
0.4$R_{25}$ (Bigiel et al.~2008), so they might have increased metallicities and
dust-to-gas ratios with depressed $X_{\rm CO}$ values, which would increase the
Eddington ratio (see Figure \ref{fig:radprofiles6946}, \S \ref{sec:DGR} and
\ref{sec:xcoxhcn}). Super-Eddington apertures contain 6\% of the total flux in
$\rm H_2$-detected apertures across the whole sample; in NGC 6946, for example,
super-Eddington apertures contain 10\% of the total flux. The super-Eddington
apertures indicate that radiation pressure can be dynamically dominant even when
individual star-forming regions remain unresolved, suggesting that radiation
pressure may be more important on the scale of GMCs and massive star
clusters. Finally, we calculated $\Gamma_{\rm H_2}$ assuming that UV photons
contribute to the radiation pressure $P_{\rm rad}=(F_{\rm UV}+F_{\rm
IR})/c$. The distribution of $\Gamma_{\rm H_2}$ remains nearly the same because
star-forming regions have $F_{\rm IR}/F_{\rm UV} \gg 1$, so we refrain from
plotting it in Figure \ref{fig:inthist}.

\section{Discussion}
\label{sec:discussion}
We have compared globally-averaged and resolved observations of star-forming
galaxies with theoretical expectations based on the theory of radiation pressure
supported star formation (see \S2).  Although the uncertainties are large (see
below), our primary findings are as follows.\\

\noindent 1.~Figures \ref{fig:lirlco}--\ref{fig:kslaw_pradpmid} show that
star-forming galaxies meet, but do not dramatically exceed, nominal expectations
for the dust Eddington limit.  When some subregions do seem to exceed the
Eddington limit (as in the outer regions of galaxies in Figure
\ref{fig:radprofiles} \& \ref{fig:radprofiles6946}), we consider this to be
consistent with Eddington since trends in the dust-to-gas ratio and CO-to-H$_2$
conversion factor, as well as the large-scale stellar potential and
intermittency of the star-formation process ($\xi$; eq.~\ref{eqn:int}) affect
the Eddington ratio at order unity.\\

\noindent 2.~The $L_{\rm IR}$--$L^\prime_{\rm HCN}$ plot (Figure
\ref{fig:lirlhcn}) provides the strongest evidence for the importance of
radiation pressure feedback since $L^\prime_{\rm HCN}$ is expected to directly
trace the dense actively star-forming gas and $L_{\rm IR}$ traces the total star
formation rate.  If radiation pressure in fact dominates feedback, we would
expect a one-to-one correspondence between these two quantities, and such a
relation is observed (see also Scoville et al.~2001; Scoville 2003).
Nevertheless, for typical values of both $\kappa$ in the optically thick limit
and the HCN-to-H$_2$ conversion factor, the Eddington limit overpredicts $L_{\rm
IR}$ by a factor of $\sim$3--6.  This discrepancy may indicate that the
dust-to-gas ratio is larger in the dense HCN-emitting regions, or that the
HCN-to-H$_2$ conversion factor is smaller (see \S \ref{sec:xcoxhcn} below).  If
radiation pressure feedback regulates star formation, then this relation is in a
sense more fundamental than the Schmidt Law because HCN-emitting gas is more
closely connected with star formation than CO-emitting gas, in which case we
would expect $\dot{\Sigma}_\star^{\rm Edd} = 4 \pi G \Sigma_{\rm H_2}^{\rm
dense}/(\epsilon c \kappa_{\rm FIR})$.\\

\noindent 3.~The central regions of all galaxies in Figure \ref{fig:radprofiles}
are prima facie substantially sub-Eddington when a constant dust-to-gas ratio
and CO-to-H$_2$ conversion factor are applied to all sub-regions without regard to
their radial location.  If radiation pressure is in fact the dominant feedback
mechanism in these regions, a much higher central dust-to-gas ratio and a lower
CO-to-H$_2$ conversion factor are required (see Figures \ref{fig:radprofiles} \&
\ref{fig:radprofiles6946}). It would be particularly useful for testing
radiation pressure feedback to produce the same profiles in HCN.\\

\noindent 4.~The ``break'' in the observed Schmidt Law at $\Sigma_{\rm
g}\sim100$--$1000$ M$_\odot$ pc$^{-2}$ (see Figure \ref{fig:kslaw_pradpmid};
Daddi et al. 2010b) may be due to the transition from the single-scattering
limit to the optically thick limit in the GMCs that collapse to form stars, as
in MQT.\\

\noindent 5.~If radiation pressure is the primary feedback mechanism for
regulating star formation, then we predict that the Schmidt Law will follow the
form of eq.~\ref{eqn:schmidtlaw} (for discussion of $\kappa$ and $\xi$ as well
as uncertainties see \S \ref{sec:theory} \&
\ref{sec:inter-disc}--\ref{sec:xcoxhcn}).\\

\noindent 6.~A testable prediction of radiation pressure feedback is that all
else being equal the star formation rate should depend linearly on the
dust-to-gas ratio in the optically thick limit.\\

In addition to these points, below we note an implication of radiation pressure
feedback that has so far not been stated in the literature (\S \ref{sec:tgas}).
Finally, in the remaining subsections we highlight the dominant uncertainties in
our work as a guide for future research on the importance of radiation pressure
feedback in star-forming galaxies.

\subsection{Gas Depletion Timescale}
\label{sec:tgas}
The gas depletion timescale, the time required to consume a galaxy's gas
reservoir at the current SFR, is observed to be $\sim$2 Gyr in normal spirals
(Kennicutt 1998; Leroy et al.~2008). Radiation pressure sets the gas depletion
timescale to be
\begin{equation}
 t_{\rm gas} = \frac{M_{\rm g}}{\dot{M}_\star} = \frac{M_{\rm g} \epsilon
 c^2}{\xi L_{\rm Edd}} = \frac{\epsilon c \kappa}{4 \pi G \xi}.
\end{equation}
Using typical numbers for a spiral galaxy in the single-scattering limit, the
gas depletion timescale is
\begin{equation}
 t_{\rm gas} \sim 2.1 \; {\rm Gyr} \; \Sigma_{30}^{-3/2} \, h_{100}^{1/2},
\label{eqn:tgasSS}
\end{equation}
where $\Sigma_{30}=\Sigma/30 \; \rm M_\odot \; pc^{-2}$ and $h_{100}=h/100 \;
\rm pc$.  This normalization of $t_{\rm gas}$ is in good agreement with the
observed gas depletion timescale, but eq.~\ref{eqn:tgasSS} predicts that the gas
depletion timescale should have a strong dependence on $\Sigma_{\rm g}$, in
contrast to the observations of Leroy et al.~(2008) (see their Figure 15).  For
completeness, we note that variations in the dust-to-gas ratio will not affect
the dependence of $t_{\rm gas}$ on $\Sigma_{\rm g}$ in the single-scattering
limit, but uncertainties in $X_{\rm CO}$, $\xi$, and $\phi$ (see \S
\ref{sec:DGR}--\ref{sec:xcoxhcn}) might impact the gas depletion timescale.

Hot starbursts and optically thick subregions (see \S \ref{sec:optthick}) have
intermittency factors that approach unity and nearly constant opacities, so the
gas depletion time is approximately constant,
\begin{equation}
 t_{\rm gas} \approx 5.7 \; {\rm Myr} \; \kappa_{10},
\end{equation}
where $\kappa_{10}=\kappa_{\rm FIR}/10 \; \rm cm^2 \; g^{-1}$. For comparison,
Sakamoto et al.~(2008) find that the optically thick western nucleus of Arp 220
has a gas depletion time of $\sim$6 Myr. The fact that the gas depletion
timescale set by radiation pressure is consistent with the observed gas
depletion timescale in spirals and ULIRGs is equivalent to the statement of
Figures \ref{fig:lirlco}--\ref{fig:radprofiles6946} that starbursts approach the
dust Eddington limit.  In addition, we point out that radiation pressure
feedback predicts that the specific SFR (SSFR) will be ${\rm SSFR} \sim 4 \pi G
\xi f_{\rm gas}/(\epsilon c \kappa)$ for small $f_{\rm gas}$.

\subsection{Intermittency}
\label{sec:inter-disc}
The intermittency factor $\xi$ (see \S \ref{sec:intermittency}; MQT) relates the
properties of radiation pressure dominated star-forming subregions to the global
properties of a galaxy. However, $\xi$ may overestimate the effect of
intermittency in some galaxies (see Figures \ref{fig:kslaw_pradpmid},
\ref{fig:radprofiles}, \& \ref{fig:inthist}). We expect the determination of
$\xi$ to be complicated by uncertainty in the timescale for the central cluster
of a sub-unit to be bright ($t_{\rm MS} \sim 4 \; \rm Myr$). We adopt $t_{\rm
MS}$ as the time a cluster will be bright, since the cluster luminosity drops
rapidly after the most massive stars in the cluster explode as
supernovae. However, a cluster will continue to emit after $t_{\rm MS}$. Indeed,
models of cluster luminosity indicate that a similar amount of momentum will
transferred to the gas during the time $t_0 \rightarrow t_{\rm MS}$ and during
the time $t_{\rm MS} \rightarrow 4t_{\rm MS}$, where the cluster luminosity has
dropped by 1 dex after $\sim$4$t_{\rm MS}$ (Leitherer et al.~1999). Further
uncertainty in $\xi$ is due to ambiguities in calculating the lifetime of a
sub-unit ($\sim$2$t_{\rm dyn}+t_{\rm MS}$), especially the disk dynamical time
(see eq.~\ref{eqn:tdyn}). The dynamical time likely varies with galactocentric
radius because $\Sigma$ is a strong function of radius (Leroy et
al.~2008). Thus, trying to determine an effective $\xi$ applicable to a galaxy
as a whole may be difficult if $\xi$ changes locally. Overall, we expect the
uncertainty in $\xi$ to be a factor of a few to several.

\subsection{The FIR Optical Depth}
\label{sec:tauFIR}
A key theoretical uncertainty in calculating the Eddington limit for dense
starbursts is the effective optical depth ($\tau_{\rm eff}$) for surface
densities where $\tau_{\rm FIR} > 1$. In order for radiation pressure to be
dynamically important in optically thick GMCs, $\tau_{\rm eff}$ must exceed
unity.  Based on the high Mach number turbulence simulations of Ostriker et
al. (2001), MQT conclude that if the ISM is optically thick on average, then the
vast majority of sight lines will be optically thick. For comparison, KM09 argue
that instabilities, such as Rayleigh-Taylor and photon-bubble instabilities,
will reduce the effective optical depth of the dense ISM to $\sim$1. However,
MQT note that both the midplane pressure from gravity $P_{\rm mid} \sim \pi G
\Sigma^2$ and optically thick radiation pressure $P_{\rm rad} \sim \tau F/c
\propto \Sigma^2$ scale as $\Sigma^2$, a feature unique to radiation pressure
among stellar feedback processes.  Thus, if radiation pressure cannot regulate
star formation in dense, optically thick gas, then no known stellar feedback
process can.

\subsection{Dust-to-Gas Ratio and Metallicity}
\label{sec:DGR}
The coupling between radiation and gas directly depends on the dust-to-gas ratio
($\kappa \propto f_{\rm dg}$). In this analysis, we assume the Galactic value
for the dust-to-gas ratio ($f_{\rm dg}=1/150$) and solar metallicity; however,
there is strong evidence that $f_{\rm dg}$ and metallicity change with
environment. The dust-to-gas ratio has been shown to correlate with metallicity
and radius (Issa et al.~1990; Lisenfeld \& Ferrara 1998; Draine et al.~2007;
Mu{\~n}oz-Mateos et al.~2009). Mu{\~n}oz-Mateos et al.~(2009) find that the
dust-to-gas ratio can climb to values as high as $f_{\rm dg} \sim 1/10$ in the
centers of spiral galaxies. This increase in metallicity and dust-to-gas ratio
is necessary for the centers of star-forming spirals to be at Eddington (see
Figures \ref{fig:radprofiles} \& \ref{fig:radprofiles6946} and \S
\ref{sec:subgalactic_scales}). The average dust-to-gas ratio of local spirals
also varies by a factor of a few (e.g., M51 has a $f_{\rm dg} \sim 1/75$; Draine
et al.~2007). Furthermore, the dust-to-gas ratio is observed to be higher in
some dense starbursts, such as SMGs ($f_{\rm dg} \sim 1/50$; Kov\'{a}cs et
al.~2006; Micha{\l}owski et al.~2010) and sub-mm faint ULIRGs ($f_{\rm dg} \sim
1/20$; Casey et al.~2009). Importantly, if we adopt a dust-to-gas ratio
potentially appropriate for dusty starbursts (short dashed line in Figure
\ref{fig:lirlco}, \ref{fig:lirlhcn}, and the left panel of Figure
\ref{fig:kslaw_pradpmid}), then a substantial fraction of optically thick
galaxies would be at or above the Eddington limit (Figure
\ref{fig:radprofiles6946} illustrates this for NGC 6946).

\subsection{Molecular Gas Tracers}
\label{sec:xcoxhcn}
The Eddington limit depends strongly on the CO-to-H$_2$ ($X_{\rm CO}$) and the
HCN-to-H$_2$ ($X_{\rm HCN}$) conversion factors (see eqs. \ref{eqn:leddlco} \&
\ref{eqn:leddlhcn}). These conversion factors are two of the largest sources of
observational uncertainty in our calculations because they vary with excitation
conditions ($X \propto \sqrt{n_{\rm H_2}}/T_{\rm b}$, where $T_{\rm b}$ is the
brightness temperature) and metallicity. Several lines of evidence suggest that
$X_{\rm CO}^{\rm MW}$ overestimates $M_{\rm H_2}$ in starburst galaxies. For
example, $X_{\rm CO}$ is a factor of $\sim$3 lower in M82 (Weiss et al.~2001)
and a factor of $\sim$5 lower in a sample of local ULIRGs (Downes \& Solomon
1998). To account for the different values of $X_{\rm CO}$ in normal spirals and
extreme starbursts, we apply the Milky Way $X_{\rm CO}$ value to galaxies with
$L_{\rm IR} < 10^{11} L_\odot$ and the ULIRG $X_{\rm CO}$ value to galaxies with
$L_{\rm IR} > 10^{11} L_\odot$. Because this prescription is somewhat
simplistic, it probably overestimates $M_{\rm H_2}$ in moderate luminosity
starbursts ($L_{\rm IR} < 10^{11} L_\odot$), such as M82, and in the centers of
star-forming spirals (see Figures \ref{fig:radprofiles} \&
\ref{fig:radprofiles6946} and \S \ref{sec:subgalactic_scales}). Additionally, it
likely underestimates $M_{\rm H_2}$ in ultra-luminous ($L_{\rm IR} \sim 10^{12}
\; \rm L_\odot$) high redshift disk (BzK) galaxies, for which Daddi et
al.~(2010a) find a value of $X_{\rm CO}$ that is consistent with the Galactic
value. As a result, moderate luminosity starbursts and the centers of
star-forming spirals may be closer to Eddington and BzK galaxies might be
further below the optically thick Eddington limit than Figure
\ref{fig:kslaw_pradpmid} would suggest.
%update with Daddi CO 1-0 data

Unfortunately, $X_{\rm HCN}$ is more uncertain than $X_{\rm CO}$ because there
is no direct calibration of $X_{\rm HCN}$ from Milky Way GMCs. For normal
spirals, Gao \& Solomon (2004a,b) find $X_{\rm HCN} \sim 10 \;\rm M_{\odot}
(K\;km\;s^{-1}\;pc^2)^{-1}$ for virialized cloud cores with $\langle n \rangle =
3\times 10^4 \; \rm cm^{-3}$ and $T_{\rm b}=35 \; \rm K$. They caution that
$X_{\rm HCN}$ could be lower in regions of massive star formation due to
significantly higher brightness temperatures $T_{\rm b} \sim {\rm few} \times
10^2 \; \rm K$ (Boonman et al.~2001). Ultra-luminous starbursts exhibit
widespread intense massive star formation, so one might expect that $X_{\rm
HCN}$ is lower in more luminous galaxies. For example, Graci{\'a}-Carpio et
al.~(2008) estimate that $X_{\rm HCN}$ should be $\sim$4.5 times lower for
galaxies at $L_{\rm FIR} \sim 10^{12} \; \rm L_{\odot}$ than at $L_{\rm FIR}
\sim 10^{11} \; \rm L_{\odot}$. We note that if $X_{\rm HCN}$ is smaller than
the assumed value of $3 \;\rm M_\odot (K\;km\;s^{-1}\;pc^2)^{-1}$, then more
galaxies will approach or exceed the Eddington limit (see Figure
\ref{fig:lirlhcn}). For example, decreasing $X_{\rm HCN}$ by a factor of $\sim
2$ brings essentially all galaxies in line with the Eddington limit for the
nominal value of $\kappa_{\rm FIR}$ ($\sim$5--10).

\acknowledgments 

We thank N. Murray, E. Quataert, P. Hopkins, P. Martini, and R. Pogge for
helpful conversations. We thank A. Leroy, F. Bigiel, L. Yan, E. Daddi, T. Greve,
and D. Riechers for providing us with their data. We also thank the referee,
Mark Krumholz, for a timely and useful report, and for emphasizing the point
made in footnote 5. T.A.T. is supported in part by an Alfred P. Sloan Foundation
Fellowship. This work is supported by NASA grant \#NNX10AD01G.

\end{document}